# On the morphodynamics of a wide class of large-scale meandering rivers: Insights gained by coupling LES with sediment-dynamics

Ali Khosronejad[1], Ajay B. Limaye[2], Zexia Zhang[1], Seokkoo Kang[3], Xiaolei Yang[4], and Fotis Sotiropoulos[5]

[1]Civil Engineering Dept., Stony Brook University, Stony Brook, NY 11794, USA

[2]Dept. of Environmental Sciences, University of Virginia, Charlottesville, VA 22904, USA

[3]Dept. of Civil and Environmental, Engineering, Hanyang University, Seoul, South Korea

[4]State Key Laboratory of Nonlinear Mechanics, Institute of Mechanics, Chinese Academy of Sciences, Beijing 100190, China

[5]Dept. of Mechanical and Nuclear Eng., Virginia Commonwealth University, Richmond, VA 23284, USA

Corresponding author's Email: ali.khosronejad@stonybrook.edu

**Abstract**

In meandering rivers, interactions between flow, sediment transport, and bed topography affect diverse processes, including bedform development and channel migration. Predicting how these interactions affect the spatial patterns and magnitudes of bed deformation in meandering rivers is essential for various river engineering and geoscience problems. Computational fluid dynamics simulations can predict river morphodynamics at fine temporal and spatial scales but have traditionally been challenged by the large scale of natural rivers. We conducted coupled large-eddy simulation (LES) and bed morphodynamics simulations to create a unique database of hydro-morphodynamic datasets for 42 meandering rivers with a variety of planform shapes and large-scale geometrical features that mimic natural meanders. For each simulated river, the database includes (*i*) bed morphology, (*ii*) three-dimensional mean velocity field, and (*iii*) bed shear stress distribution under bankfull flow conditions. The calculated morphodynamics results at dynamic equilibrium revealed the formation of scour and deposition patterns near the outer and inner banks, respectively, while the location of point bars and scour regions around the apexes of the meander bends is found to vary as a function of the radius of curvature of the bends to the width ratio. A new mechanism is proposed that explains this seemingly paradoxical finding. The high-fidelity simulation results generated in this work provide researchers and scientists with a rich numerical database for morphodynamics and bed shear stress distributions in large-scale meandering rivers to enable systematic investigation of the underlying phenomena and support a range of river engineering applications.

## 1. Introduction

Rivers are the focal points of human society and Earth systems. Rivers are vital corridors for shipping, manufacturing, agriculture, and recreation. Water-borne pollutants, as well as those carried in sediments, are deposited during floods, impact surfaces and groundwater quality. Rivers are the primary habitat for aquatic species and provide an indispensable ecosystem for riparian species. Rivers transport enormous quantities of sediments and solutes from continents to the ocean (Syvitski and Milliman, 2007), shaping continental surfaces and coastlines, controlling the stacking and connectivity of subsurface fluid reservoirs (Gibling, 2006), and influencing global geochemical cycles (Berner, 1982). A study of river morphodynamics is important from the point of view of engineering and design, as floods can cause damage to infrastructure installed in rivers. Over 30% of bridge failures in the United States are caused by scour during floods (e.g., Wardhana & Hadipriono, 2003; Biezma & Schanack, 2007; Hughes et al., 2007; Deng et al., 2015). Major challenges in developing sustainable strategies for assessing and mitigating flood damage arise from intricate interactions between river morphology and turbulent flood flow dynamics. Such interactions, which are site-specific and occur across a broad range of scales, must be adequately predicted to assess the efficacy of potential flood mitigation strategies.

Given this variety of applications, it is necessary to predict how the flow, mobile sediment, and topography interact in natural channels. However, a key complication is the sinuous shapes of single-thread natural channels, where the flow travels through bends of varying shapes and amplitudes (Fig. 1). The patterns of flow in river bends have long been a focus of study (Thomson, 1877; Rozovski, 1957). In a classic model, the centrifugal force and associated pressure gradients that are induced by the curvature of the bend cause a spiral flow pattern with the surface flow deflected toward the outer bank and the low momentum bed flow returning back toward the inner bank. The interaction between this complex secondary flow pattern with the primary flow and bed topography contributes to complex morphodynamic phenomena, including bank erosion and point bar deposition (Dietrich, 1987). Subsequent studies have shown that, in addition to these effects, turbulence anisotropy can cause additional longitudinal flow structures in meander bends that rotate in the opposite direction to the primary centrifugally-induced main cell. Such structures can influence the magnitude and distribution of the shear stress along the channel and sediment transport and affect erosion and deposition patterns (e.g., Kang & Sotiropoulos, 2012; Papanicolaou et al., 2007; Keylock et al., 2005).

Increasingly, computational fluid dynamics models have been used to investigate three-dimensional turbulent flow in sinuous channels (Shimizu et al., 1992; Ye and McCorquodale, 1998; Blanckaert and DeVriend, 2003; Ferguson et al., 2003; Wilson et al., 2003; Ruther and Olsen, 2005; Rodriguez et al., 2005). Nonetheless, simulating morphodynamic processes in rivers under flood conditions poses major computational challenges for high-fidelity numerical models (Keylock et al., 2005; Camporeale et al., 2007; Stoesser, 2014). The computational domains are large, with typical river widths spanning tens to hundreds of meters and lengths spanning kilometers. The curving planform geometry and irregular cross-sections of meandering channels complicate grid generation, and the feedback between turbulent flow, sediment transport, and the topography of the channel bed requires fine spatial and temporal resolution to analyze. Finally, there is a large disparity between the time scale of river flow dynamics versus the time scale of sediment dynamics which leads to bed deformations (Khosronejad et al., 2014). As a result, previous studies using high-fidelity flow models, including large-eddy simulations and direct numerical simulations, have mostly focused on highly simplified geometries, e.g., 90° bends (Ruther and Olsen, 2005). To circumvent the high cost of high-fidelity modeling, researchers have recently attempted to develop machine-learning algorithms to predict flood flow fields (Mosavi et al., 2018; Qian et al., 2019; Zhang et al., 2022). Despite their potential, such machine-learning approaches are in their infancy and require more research to enable the prediction of flood-induced flow fields and bed deformations in large-scale rivers. Moreover, such methods need extensive datasets for a variety of meandering rivers for training purposes. For example, Zhang et al. (2022) employed the flow field of a large-scale meandering river, produced using large-eddy simulation (LES), to develop a machine-learning algorithm that allowed for the prediction of flood-induced flow fields in other large-scale meandering rivers. Their study was limited to flood flow predictions without the ability to generate the equilibrium bed morphology of the rivers.

This study aims to develop the first database of high-fidelity morphodynamic simulations for a wide range of meandering channel geometries representative in shape and scale of those encountered in nature. We developed simulations of 3D flow fields under bed morphology and bed shear stress at flood discharges. To reduce the high cost of this undertaking, we leveraged recent advances in parallel computing and numerical algorithms that rendered our computational efforts more efficient. The modeling infrastructure supports field-scale river bends (i.e., kilometer-scale domains) with realistic, evolving bottom topography (Khosronejad and Sotiropolous, 2014), complex geometries, and sufficiently high Reynolds numbers to capture well-developed turbulent flows (Kang and Soritopoulos, 2012).

We employed our in-house high-fidelity model, the so-called Virtual Flow Simulator (VFS-Geophysics) code, to generate a database of the flood-induced flow field, bed morphology, and bed shear stress distribution of 42 large-scale meandering rivers. The planform shapes, hydraulic conditions, and sediment properties of these rivers were selected to represent a wide range of naturally occurring rivers and flood conditions. Numerical simulations were conducted by coupling LES with a sediment transport model that accounts for motion as both bed load and suspended load. The computed bed morphology of the rivers highlights the formation and migration of bed forms. The simulation results of this study provide a novel database for the riverbed deformation and bed shear stress distribution for a wide variety of large-scale meandering rivers, which can be used by researchers and scientists to help (*i*) better understand the relationship between channel planform geometry and bed morphology, and (*ii*) assess and mitigate flood events in similar natural environments.

The remainder of this paper is organized as follows. In Section 2, we present the governing equations for the flow and sediment transport. In Section 3, we describe common river shapes and related geometric models to motivate the design of 42 large-scale meandering rivers as testbeds for this study. The computational details of the simulations are presented in Section 4. The simulation results are presented in Section 5, followed by conclusions in Section 6.

**2. Governing equations**

In this section, we briefly outline the equations governing the hydrodynamics and morphodynamics of mobile-bed river flows in our modeling approach. For a detailed description of the mathematical formulation of the model, the hydro- and morphodynamic coupling technique used in this study can be found elsewhere (see Yang et al., 2017; Khosronejad & Sotiropoulos, 2014; and Khosronejad et al., 2011, 2012, 2013, 2014, 2015, 2019a,b, 2020c).

## 2.1. Hydrodynamics

To quantify the bankfull river flow, we solved the spatially averaged continuity and Navier-Stokes equations, which govern the instantaneous, resolved flow field for 3D, incompressible, turbulent river flows. The governing equations of Cartesian coordinates, $\{x_i\}$, are transformed to curvilinear coordinates, $\{\xi_i\}$, in which the compact tensor notation form of the equations is as follows ($i = 1, 2,$ or $3$, and $j, k, l = 1, 2,$ and $3$):

$$J \frac{\partial U_j}{\partial \xi_j} = 0 \tag{1}$$

$$\frac{1}{J}\frac{\partial U_i}{\partial t} = \frac{\xi_l^i}{J}\left[-\frac{\partial(U_j u_l)}{\partial \xi_j} + \frac{\partial}{\partial \xi_j}\left(\nu \frac{G^{jk}}{J}\frac{\partial u_l}{\partial \xi_k}\right) - \frac{1}{\rho}\frac{\partial}{\partial \xi_j}\left(\frac{\xi_l^i p}{J}\right) - \frac{1}{\rho}\frac{\partial \tau_{lj}}{\partial \xi_j}\right] \tag{2}$$

where $J$ is the Jacobian of the geometric transformation from Cartesian to curvilinear coordinates, $\xi_l^i = \partial \xi_i / \partial x_l$ are the transformation metrics, $u_i$ is the $i^{th}$ Cartesian velocity component, $U_i = (\xi_l^i/J)u_l$ is the contravariant volume flux, $\nu$ is the kinematic viscosity of water, $G^{jk} = \xi_l^i \xi_l^k$ are the components of the contravariant metric tensor, $p$ is the pressure, $\rho$ is the density of water, and $\tau_{lj}$ is the subgrid stress tensor of the LES. Here, we highlight the key aspects of our LES method, which was implemented in the context of the curvilinear immersed boundary (CURVIB) method to allow for the simulation of arbitrarily complex geometric configurations such as those of meandering rivers with deformable mobile beds. The filtered Navier-Stokes equations were obtained by decomposing the velocity field into resolved and unresolved components and integrating the equations over the spatial filter. Consequently, the sub-grid stress terms appear in the momentum equations. To model these stress terms, we employed the dynamic Smagorinsky sub-grid scale (SGS) model as follows:

$$\tau_{ij} - \frac{\delta_{ij}}{3}\tau_{kk} = -2\mu_t \overline{S_{ij}} \tag{3}$$

where $\delta_{ij}$ is the Kronecker delta function, $\overline{S_{ij}}$ is the filtered strain-rate tensor in which the overbar denotes the grid filtering operation, $\mu_t$ is the eddy viscosity defined as $\mu_t = C_s \Delta^2 |\bar{S}|$, $C_s$ is the Smagorinsky contact, $\Delta = J^{-1/3}$ is the filter size calculated by the box filter, and $|\bar{S}| = \sqrt{2\overline{S_{ij}}\,\overline{S_{ij}}}$. Using the dynamic Smagorinsky SGS model, constant $C_s$ evolves in time and space as a function

of the flow field and is more appropriate for turbulent flows with high Reynolds numbers, such as those in natural river channels (for more details see Kang et al., 2011).

The governing equations were discretized in space on a hybrid staggered/non-staggered computational grid arrangement using a central, second-order accurate numerical scheme for the convective, divergence, pressure gradient, and viscous-like terms. For the time derivatives, second-order backward differencing was used, and the equations were integrated in time using a second-order accurate fractional step methodology. A Jacobian-free Newton-Krylov solver was used in conjunction with the fractional step method to handle momentum equations. Finally, we employed a generalized minimal residual method-based solver, enhanced with a multigrid as a preconditioner, to solve the Poisson equation. For more details regarding the numerical methods used to discretize the governing equations in time and space, see Kang et al. (2011).

It is important to note that transient bed deformation on the mobile beds of rivers can considerably complicate the generation of good quality boundary-fitted grid systems. Therefore, we utilized the CURVIB method, which allows for efficient flow and sediment dynamics simulations in arbitrarily complex geometries with irregular, arbitrarily large bed deformations [REF]. In the context of the CURVIB method, the background computational domain for each meandering river exactly follows the curvature of the river and is discretized with a curvilinear grid system. The riverbanks and sediment/water interface, which are immersed in the background grid system, are discretized with unstructured triangular grids. The governing equations for hydrodynamics were solved at the background grid nodes in the fluid phase with boundary conditions specified at fluid nodes in the immediate vicinity of the sediment/water and riverbanks. The boundary nodes are referred to as the immersed boundary (IB) nodes. The computational nodes inside the unstructured triangular grid system (i.e., the riverbanks and sediment layer) were removed from computations. The flow field boundary conditions were reconstructed at the IB nodes using the wall modeling approach developed for the CURVIB framework (Gilmanov & Sotiropoulos, 2005, Kang et al., 2011). At the end of each time step, the riverbed deforms and, thus, we conducted a new search to categorize the computational nodes as fluid, i.e., IB, or solid, i.e., nodes inside the riverbanks and the sediment layer. To re-classify the grid nodes as the bed geometry changes owing to local scour or large-scale sediment transport along the river, we used an efficient ray-tracing algorithm (for more detail see Borazjani et al., 2008).

**2.2. Morphodynamics**

In this study, we considered the bed deformations due to both the bed and suspended sediment loads occurring within the thin bed-load layer and the flow domain, respectively. As a result, the temporal variation in bed elevation is governed by the non-equilibrium equation for sediment mass balance, also known as the Exner-Polya equation:

$$(1 - \varphi)\frac{\partial z_b}{\partial t} = -\nabla \cdot \mathbf{q}_{BL} + D_b - E_b \tag{4}$$

where $\varphi$ is the sediment martial porosity (= 0.4), $z_b$ is the bed elevation, $\mathbf{q}_{BL}$ is the bed-load flux vector, $\nabla$ is the divergence operator, $D_b$ is the rate of net sediment deposition from the suspension onto the bed, and $E_b$ is the rate of net sediment entrainment from the bed into the flow domain, also known as the particle pick-up rate. The bed load flux vector within the bed load layer is obtained as follows:

$$\mathbf{q}_{BL} = C_{BL}\delta_{BL}\mathbf{u}_{BL} \tag{4}$$

where $C_{BL}$ is the sediment concentration within the bed load layer with a thickness of $\delta_{BL}$, and $\mathbf{u}_{BL}$ is the velocity vector parallel to the bed surface at the interface of water and sediment, that is, at the top of the bed load layer. The bed-load sediment concentration $C_{BL}$ and bed-load layer thickness $\delta_{BL}$ were calculated using van Rijn (1993) formulations as functions of the local bed shear stress and the threshold of motion of sediment particles. The latter was obtained using the Shields criterion for a flat bed and then corrected for the local transverse and/or longitudinal bed slopes (Shields, 1936). All parameters in this equation were calculated instantaneously for each local point over the mobile beds of the rivers.

The net rates of sediment deposition $D_b$ and entrainment $E_b$ over the mobile bed are computed as

$$D_b = w_s C_b \tag{5}$$

$$E_b = w_s C_{BL} \tag{6}$$

where $w_s$ is the settling velocity of the non-spherical sediment particles obtained from van Rijn's formula (see van Rijn, 1993), and $C_b$ is the instantaneous sediment concentration immediately above the bed load layer. The deposition rate is related to the sediment material, which is vertically transported from the flow domain onto the mobile bed. Therefore, we employed a quadratic interpolation method to calculate $C_b$ from the concentration field $C$ of the suspended sediment within the flow domain.

For a dilute sediment-water mixture, in which the volumetric sediment concentration is less than O(0.01), the concentration field of the suspended sediment is modeled as a passive tracer using the following convection-diffusion equation:

$$\frac{1}{J}\frac{\partial(\rho C)}{\partial t} + \frac{\partial\left(\rho C\left(U_j - W_j \delta_{ij}\right)\right)}{\partial \xi_j} = \frac{\partial}{\partial \xi_j}\left(\left(\frac{\nu}{S_L} + \frac{\nu_t}{S_T}\right)\frac{G^{jk}}{J}\frac{\partial C}{\partial \xi_k}\right) \quad (7)$$

where $W_j = (\xi_3^j/J)w_s$ is the vertical contravariant volume flux of the suspended sediment concentration owing to the settling velocity of particles within the flow domain, $S_L$ is the laminar Schmidt number (= 700), $S_T$ is the turbulent Schmidt number (= 0.75), and $\nu_t$ is the kinematic eddy viscosity. The convection-diffusion equation is discretized using the second-order central differencing numerical scheme and solved using the fully implicit Jacobian free Newton method.

Finally, to prevent the computed bed slope from exceeding the angle of repose of the sediment material, we employed a mass-conservative sand slide model. Once the bed elevations are computed, the sand-slide algorithm is activated to ensure that the bed slope at each computational bed surface cell is less than or equal to the angle of repose. The sand-slide algorithm corrects the slope of cells with bed slopes greater than the angle of repose by redistributing the excess sediment material over other neighboring cells until all slopes are less than or equal to the angle of repose (for more details see Khosronejad et al., 2011, Khosronejad & Sotiropoulos (2014, 2017)).

### 2.3. Coupling of hydrodynamics and morphodynamics

The coupling between the turbulent flow and morphodynamics simulations was performed using the partitioned loose-coupling fluid-structure interaction (FSI) method (Khosronejad et al., 2011). In this approach, we solve the governing equations of the turbulent flow and morphodynamics separately in the water and sediment domains while accounting for the interaction of the two phases by applying the boundary conditions at the interface of the water and sediment, that is, at the top of the bed load layer. When the flow field equations are solved we consider boundary conditions specified over the mobile bed in terms of its location and geometry, the rate of bed elevation change (i.e., the vertical velocity of the bed surface), and the concentration of the sediment at the top of the bed load layer. However, when solving the morphodynamics equations, the near-bed velocity field, bed shear stress, and concentration of suspended sediment near the water-sediment interface are utilized to compute the sediment flux within the bed-load layer and net vertical sediment flux onto the bed surface.

Using the loose-coupling FSI approach for flow-bed interactions with the assumption that the bed surface geometry, rate of change of the bed surface, and concentration fields of flow and suspended sediment are available at time step $n$, we use the following algorithm to solve for the flow field and bed morphodynamics at time step $n + 1$ (Khosronejad & Sotiropoulos, 2014):

i) Compute the river flow field at time $n + 1$ by solving the hydrodynamic governing equations using the known bed geometry and bed change rate of time $n$;

ii) compute the suspended sediment concentration field at time $n + 1$ by solving the convection-diffusion equation of sediment material using the known bed geometry and flow field at times $n$ and $n + 1$, respectively.

iii) Calculate the bed changes and, subsequently, the new bed geometry at time $n + 1$ by solving the morphodynamic equations using the known flow field and suspended sediment concentration field at time $n + 1$;

iv) compute the rate of bed elevation change to provide the boundary conditions for the flow field simulations of the next time step.

Finally, a key aspect of the simulations is that we employ different time steps to match the hydrodynamic and morphodynamic modules of the code with respect to time. Given that, for most applications in geophysical flows, the characteristic time scale of the morphodynamics is an order of magnitude larger than that of the hydrodynamics (Mercier et al. 2012), different time steps for the flow and morphodynamics alleviate the high computational costs of the coupled simulations. More specifically, for river simulations, the coupled system of flow and morphodynamics should be run long enough to obtain the dynamic equilibrium bed morphology covering the time scale of the morphodynamic evolution. However, small time steps (comparable to the time scale of resolved eddies in the LES) are needed to resolve the turbulent flow field. Adopting the LES time step as a common time step for both hydrodynamics and morphodynamic simulations, however, can significantly increase the overall computational cost of coupled morpho- and hydrodynamics simulations and without adding any additional resolution to the simulation of large-scale morphodynamics, which evolve at much slower scales. For example, for the flow and mobile bed interaction case in this study, the time scale of the flow obtained by the ratio of the length scale (i.e., the mean flow depth of the riverine flow) to the velocity scale (i.e., the bulk velocity of the riverine flow) ranges between about 0.7 s to 4 s. In contrast, the typical time scale for mobile bed evolution to reach dynamic equilibrium is several months. Therefore, to mitigate the computational challenges arising from the disparate time scales of the flow and morphodynamic phases, we employ a dual-time-stepping technique along with a quasi-synchronization approach (Khosronejad et al., 2014).

Using the quasi-synchronization approach, the morphodynamics calculations employ a time step $\Delta t_m$, which is an order of magnitude larger than the time step $\Delta t_f$, used for the flow solver. For example, considering the simplicity of the explicit Euler numerical scheme for matching the timescales, the bed morphology at time $n + 1$ is obtained as follows:

$$z_b^{n+1} = z_b^n + \Delta t_m \, \text{RHS}_m^n(\mathbf{u}^n) \tag{8}$$

where $\text{RHS}_m^n$ is the right-hand-side of the Exner-Poyla equation calculated using the velocity field and the bed shear stress provided by the flow solver at time step $n$. Using this computed bed geometry, the flow solver then obtains the flow field at time step $n + 1$, as:

$$\mathbf{u}^{n+1} = \mathbf{u}^n + \Delta t_f \, \mathbf{RHS}_f^n(\mathbf{u}^n, p^n, z_b^{n+1}, w_b^{n+1}) \tag{9}$$

where $\mathbf{RHS}_f^n$ is the right-hand side of the Navier-Stokes equation computed using the bed geometry $z_b^{n+1}$, and the vertical velocity of the bed change ($w_b^{n+1} = [z_b^{n+1} - z_b^n]/\Delta t_f$) is specified at the interface of the flow and bed-load layer as the boundary condition for the flow solver. We note that depending on the ratio $\Delta t_m/\Delta t_f$, morphodynamic phenomena such as the formation and migration of bed forms may or may not be captured (for more details see Khosronejad et al., 2014). The simulation results of our coupled flow and morphodynamics model with the dual-time-stepping technique were extensively validated against laboratory- and field-scale measurements, as discussed by Kang et al. (2011), Khosronejad et al. (2012, 2013, 2014, 2015, 2016a,b,c, 2018, 2019a,b, 2020a,b,c,d,e), Le et al. (2018), Khosronejad & Sotiropoulos (2017), Flora & Khosronejad (2021a,b), and Flora et al. (2021).

## 3. Design of testbeds for meandering rivers

Numerical simulations are carried out for 42 different meandering river geometries designed to incorporate geometrical characteristics of a wide range of meandering rivers encountered in nature. In this section we review the existing models and observations of river geometry that inform the design and procedures for assembling the selected meandering river testbeds.

## 3.1. Relationships between meandering river geometry and morphodynamics

Natural, single-thread rivers develop meanders with different shapes, which have been highlighted in studies of channel migration and hydrodynamics (Table 1). Brice (1974) developed an influential framework that links observations of meandering river geometry to an evolutionary sequence driven by channel migration. High-amplitude meander bends develop several characteristic shapes that are either symmetric or asymmetric, with a skew in the upstream (Fig. 1a) or downstream (Fig. 1b) directions (Carson & Lapointe, 1983; Guo et al., 2019). Channel bends have also been distinguished based on the number of lobes, which are defined as distinct and continuous arcs of locally high curvature. In this scheme, bends with a single lobe are classified as simple, whereas bends with multiple lobes are called compounds (Fig. 1c; Brice, 1974; Frothingham & Rhoads, 2003). Resistant valley walls can restrict the development of meander bends, forcing confined bends with sharp turns and moderate amplitudes (Fig. 1d; Howard, 1996; Limaye & Lamb, 2014; Nicoll & Hickin, 2010).

The shapes of meander bends have been suggested to control the structure of turbulent riverine flow. For example, Abad and Garcia (2009a, 2009b) conducted laboratory experiments in a flume with high-amplitude, asymmetric meanders and, in those tests, reversing the flow direction was equivalent to reversing the sense asymmetry from upstream-skewed to downstream-skewed. These different orientations produced different morphodynamic consequences; the upstream-skewed condition caused weaker secondary circulation, lower flow resistance, development of bars further upstream, and shallower scour in the channel bed (Abad & Garcia, 2009a, 2009b). Güneralp and Rhoads (2009) analyzed lateral migration trajectories for natural meandering channels and found that the success of model hindcasts depended on both the shape of the meander bend and the intricacy of the morphodynamic model describing coupled feedback between flow, sediment transport, and topography, implying more complicated interactions for compound bends. Additional theoretical studies suggest that common models of meandering river migration based on a linear relationship between channel curvature and excess shear stress at the channel bank break down under certain geometries. Bends with high curvatures or relatively complex compound shapes pose the greatest challenges (Blanckaert, 2011; Camporeale et al., 2007; Güneralp & Marston, 2012). For example, theoretical and laboratory studies indicate that cross-stream secondary flow is weakened in high-curvature bends compared with predictions from linear models (Ottevanger et al., 2013).

Additional numerical and laboratory studies suggest that systematic changes in the channel planform and cross-sectional geometry influence morphodynamics. Kashyap et al. (2012), for example, used a Reynolds-averaged Navier-Stokes (RANS)-based model to investigate the hydrodynamic effects of varying the channel curvature ($R/w$, where $R$ is the radius of curvature and $w$ is the channel width) and channel aspect ratio ($w/h$, where $h$ is the channel depth). These numerical experiments indicated that for the parameter ranges examined ($R/w$ = 1.5 to 10; $w/h$ = 5 to 12.5), decreases in either parameter caused systematic changes, including strong increases in cross-stream circulation and bed shear stress. Laboratory experiments by Blanckaert (2011) further indicated that secondary circulation, coherent outer-bank flow cells, and flow separation at the inner bank can impact sediment transport and stress distributions in tight bends.

Importantly, it is relatively rare for controlled laboratory or numerical studies to fully mimic natural meander bends with spatially varying curvatures, multiple bends, and asymmetric channel cross-sections formed by mobile sediment (Abad & Garcia, 2009a, 2009b). More common settings for controlled experiments involve combinations of single, symmetric bends (e.g., Blanckaert, 2011); channels with rectangular cross-sections; and the absence of sediment transport (e.g., Kashyap et al., 2012). However, numerical approaches have demonstrated promise in achieving geometric flexibility. For example, Randle (2014) constructed 72 synthetic meandering river reaches using sine curves, each with three to five symmetric bends. These simulations focused exclusively on flow properties using RANS and trapezoidal cross-sections with no sediment transport.

In summary, previous studies have identified the following key geometric characteristics for the morphodynamics of meandering channels (Table 1): bend type (simple vs. compound), planform symmetry (symmetric vs. asymmetric, with either upstream or downstream skew), radius of curvature ($R/w$), and aspect ratio ($w/h$). To date, no study has systematically varied these parameters in simulations using a mobile sediment bed.

### 3.2. Modeling channel planform geometry

Sine curves have long been recognized as simple models for highly sinuous meanders (Blanckaert, 2011; Leopold & Wolman, 1960). The Kinoshita curve mimics the centerline of a meandering river using a trigonometric function with additional terms that enable more intricate shapes (Parker et al., 1983). The local direction of the centerline is given by

$$\psi(s) = \psi_0(\sin\left(\frac{2\pi s}{\lambda}\right) + \psi_0^3(J_s \cos\left(\frac{6\pi s}{\lambda}\right) - J_f \sin\left(\frac{6\pi s}{\lambda}\right)) \tag{10}$$

where $\psi$ is the local direction of the channel centerline, $s$ is the position along the centerline, $\lambda$ is the arc length of the meander bend, $\psi_0$ is the peak angular amplitude, and $J_s$ and $J_f$ are the coefficients that control bend skewness and flatness, respectively. The Kinoshita curve has been used to design a flume to mimic a meandering channel (Abad & Garcia, 2009a, 2009b; Fernández et al., 2021) and as the initial condition for numerical experiments (e.g., Perucca et al., 2006). Computational fluid dynamic simulations with the same bend repeated three times in succession were necessary to generate fully developed turbulence and create periodic flow conditions for the middle bend (Abad & García, 2005).

### 3.3. Construction of the simulated meandering river tesbeds

Building on these previous approaches, we designed a set of numerical testbeds that isolated the key geometric properties of meanders, as summarized in Table 1. In total, we generated 42 testbeds for morphodynamics simulations (Fig. 2). Table 2 summarizes the channel geometry and mean flow properties of each testbed. We systematically varied the bend shape, radius of curvature relative to channel width ($R/w$), channel aspect ratio ($w/h$), relative roughness ($h/D$), and order of meander bend shapes along the channel, while keeping the other parameters fixed. All channels were constructed at scales typical of natural rivers. For most simulations, the channel width was fixed at 100 m, and the channel depth was fixed at 3.3 m, yielding a width-to-depth ratio of approximately 30, which is common in single-thread channels (Parker, 1976). The range for each parameter was as follows:

The first group of river testbeds varied the characteristics of the channel planform shape, all of which were generated using the Kinoshita model (Eq. 10). Rivers 1–5 were simple bends. Rivers 1, 2, and 3 represent simple symmetric bends with low, medium, and high amplitudes, respectively. River 4 is a high-amplitude, asymmetric bend with a downstream skew, and River 5 mirrors the geometry of River 4 with a downstream skew. Rivers 6–11 had compound bends. River 6 is a compound with a symmetric bend; Rivers 7 and 8 are similar but have progressively longer reaches between the main lobes. The next two rivers have compound and asymmetric bends with either an upstream skew (River 9) or a downstream skew (River 10). River 11 is also asymmetric and mimics Brice's "P" type (1974).

Rivers 12–17 are single, symmetric bends that systematically vary the minimum radius of curvature relative to the channel width ($R/w$) from 0.5 to 5 (Table 2). This parameter range was modeled after common values in the seminal study by Hickin and Nanson (1975), which linked migration rates to channel curvature on the Beatton River. The tightest bends are designed to capture the portion of the parameter space where typical low-curvature assumptions for meander morphodynamics break down (Blanckaert, 2011; Camporeale et al., 2007). River 18 is a confined meander that mimics the geometry of the bend on the Beaver River in Alberta, Canada (Fig. 1d; Parker et al., 1983; Hickin & Nicoll, 2010).

Further testbeds varied the channel cross-section geometry with respect to the base case (River 4). For Rivers 19 and 20, the channel aspect ratio ($w/h$) was varied from 10 to 100 by varying the channel width and fixing the channel depth. For Rivers 21 and 22, the relative roughness ($h/D_{50}$, where $D_{50}$ is the median grain size of the bed material) was varied compared to the base case by simultaneously increasing or decreasing both the width and depth, keeping the channel aspect ratio fixed. Relative roughness $h/D_{50}$ varied from 2000 to 20,000.

The remaining 20 testbeds were each composed of three consecutive meander bends. These more complicated geometries were designed to mimic morphodynamic conditions in natural rivers, where bends do not occur in isolation but rather encounter boundary conditions set by neighboring bends upstream and downstream. The shapes of these combinations were drawn from a subset of simple and compound bends (Rivers 2–5, and 7), the confined case (River 18), and intervening straight reaches. Table 3 lists the order of each bend sequence in Rivers 23–42.

The coupled hydro-morphodynamic simulations for all cases were initialized with a flat channel bed, such that channel cross-sections generally developed asymmetry as the simulation time advanced owing to sediment transport. In this way, the bed topography was not assumed a priory but rather computed as a result of flow and sediment transport interaction in each meandering river. The so resulting bathymetry testbeds, with spatially variable curvatures that mimic meandering rivers, comprise, to our knowledge, the most extensive set to date derived from high-fidelity, fully coupled simulations of turbulent flow and sediment dynamics. The next section provides further details on the model implementation of these testbeds.

## 4. Computational details

Each simulated meandering river bed is discretized with unstructured triangular grid systems and embedded in the flow domains, as required by the CURVIB method (Fig. 3). This figure shows that the background grid outlines the meandering river and has a depth that is sufficiently large to contain the sediment-water interface at all times. The structured mesh of the background grid system (shown on the free surface) discretizes the flow domain (Fig. 3b). Figure 3c illustrates how the unstructured triangular grid system discretizes riverbeds and banks. The use of separate grid systems in the CURVIB approach enables the handling of the arbitrary geometries of the rivers and the reconstruction of the boundary conditions at the interface between the water and the mobile sediment bed.

Table 4 shows the grid resolutions and time steps used to simulate the turbulent flow and morphodynamics of the testbed rivers. For each background grid system in this table, the grid nodes were spaced uniformly along the streamwise, spanwise, and vertical directions. In addition, unstructured triangular grid systems were uniformly spaced along the channel bed. Because the length of the testbed river is much greater than the river width and flow depth, the background grid resolution in the longitudinal direction was somewhat coarser than in the spanwise and vertical directions. We note that the present simulations correspond to high-Reynolds turbulent flow ($> 10^6$), and thus neither grid system has a resolution fine enough to resolve the viscous sublayer near the mobile sand bed and side walls. However, these grid systems, which were selected based on a series of grid sensitivity analyses, are adequate for resolving large-scale energetic coherent structures induced by (*i*) the planform geometry of the meanders and (*ii*) the deformed geometry of their beds.

The time step of the flow-field computations, $\Delta t_f$, was selected to be sufficiently small to ensure that the Courant (Friedrichs) number was less than 1.0. The time step for the morphodynamic calculations was set to $\Delta t_m = 500 \, \Delta t_f$. The use of such relatively large time steps for morphodynamics calculations within the context of our dual time-stepping desynchronization approach allowed for computationally affordable two-phase flow (of water and sediment) computations in this study. To avoid numerical stability issues, we considered several criteria when applying the dual time-stepping method. The most important criterion included imposing a limit on $\Delta t_m$ to ensure that it was sufficiently small to avoid numerical instability.

More specifically, the ratio of the flow solver and morphodynamics time steps, $\Delta t_m/\Delta t_f$, which preferably varies from 1 to 1000, was bound to avoid a phase-change event at all times during the coupled flow and morphodynamics simulations. Specific to the CURVIB immersed boundary method we employ, phase change is a detachment phenomenon during which a solid node transitions into a fluid node without experiencing an immersed boundary (IB) status, which is met when a computational node within the flow domain is placed next to the surface of the immersed boundaries. A solid node is a computational node located inside the solid bodies of sediment or sidewalls, whereas fluid nodes are located inside the flow domain. Finally, phase change occurs only in coupled flow and mobile bed simulations during which the bed elevation, and thus the geometry of the immersed body (e.g., river bathymetry), are constantly evolving (see Khosronejad & Sotiropoulos (2014) for more details).

The boundary conditions of the numerical simulations are summarized as follows. At the inlet of each computational domain, we prescribed fully turbulent open channel flows, which were calculated by performing separate precursor hydrodynamic simulations with rigid channel beds, (see Khosronejad et al., 2020, for more details). The precursor simulations were carried out in channels created by extending the inlet cross-section of the testbed rivers straight upstream by $2w$. The precursor simulations used periodic boundary conditions in the streamwise direction and were executed until the flow-field computations converged. Once converged, we stored a sufficiently long sample of the instantaneous flow field over a representative cross-plane and used it as the inflow boundary condition at the inlet cross-plane of the testbed rivers. Sediment bed loads were circulated in the testbed rivers. In other words, at each time step, we calculated the flux of the bed load sediment at the outlet. The calculated bed-load fluxes were then imposed on the next time step as the influx of bed-load sediment at the inlet. At the outlet of the computational domain, the Newman outlet boundary condition was employed for the flow-field variables and suspended sediment concentration. The free surface of the flow was treated as a sloping rigid lid with the slopes presented in Table 2.

Finally, simulations of the testbeds were carried out using 30 to 420 processors on two separate Linux clusters with over 6,000 CPUs (Intel Xeon 3.3 GHz). In other words, for the testbed rivers with the highest and lowest numbers of grid nodes (Table 4), we used 30 and 420 processors, respectively, to complete the simulations. Before starting the coupled flow and morphodynamics simulations, we first carried out precursor simulations for at least two flow-through times to obtain a statistically converged flow field for each testbed river over its flat mobile bed. The flow-through time is the amount of time required for a water particle to travel from the inlet to the outlet. On average, the run times of these precursor simulations were approximately 720–10,000 CPU hours for the testbeds with the highest and lowest number of grid nodes, respectively. For the coupled flow and morphodynamics simulations, however, the model was executed until the mobile beds of the rivers reached dynamic equilibrium. Achieving this state required approximately 11,000–450,000 CPU hours for the testbed rivers with the highest and lowest number of grid nodes, respectively. Overall, using the available computational resources, we completed coupled simulations of the 42 testbed rivers over approximately 10 months.

## 5. Results and discussion

In this section, we present the simulation results of the coupled flow and morphodynamics in the 42 testbed rivers and discuss the observed bed deformation and shear stress patterns in these rivers at dynamic equilibrium. Because this study concerns the bed morphodynamics of the testbed rivers, we herein focus on bed deformations, bed shear stress, and riverine flow fields near the mobile bed of the rivers. For detailed discussions concerning the dynamically rich vortical flow structures that form in the water column of meandering beds and streams with rigid beds, see Van Balen et al. (2009, 20210a, b), Stoesser et al. (2010), Constantinescu et al. (2011), Kang et al. (2011), Kang and Sotiropoulos (2011), and Kang and Sotiropoulos (2012a, b, c).

### 5.1. Dynamic equilibrium bed topography of the testbed rivers

The coupled flow and morphodynamics simulations of the testbed rivers under bankfull flow conditions were run until the maximum local bed change for ten successive morphodynamics time steps was less than one percent of the mean flow depth. Once this condition was reached, the simulation was stopped, and the bed morphology of the river was assumed to be in dynamic equilibrium. Figure 4 shows the bed morphology of the 42 testbed rivers in dynamic equilibrium. The bed topography data files for the testbed rivers in this figure can be downloaded from the Zenodo repository using the link provided in the data availability statement. As shown in this figure, the color map marks the bed elevation, normalized with the mean flow depth, in relation to the initial flat bed of each river. Red and blue regions indicate deposition and scour, respectively. The maximum scour depth and deposition height of the testbed rivers were predicted to be 25%–30% of their corresponding mean flow depths. The mean-flow depths of all testbed rivers, except Rivers 21 and 22, are 3.3 m, and therefore, the maximum scour depths and the height of point bars in these testbed rivers at dynamic equilibrium were predicted to be approximately 0.8 to 1 m. For Rivers 21 and 22, with mean-flow depths of 1 m and 10 m, these maxima were predicted to be approximately 0.4 m and 2.2 m, respectively.

We note that the scour/deposition patterns around the meander bend apexes are key factors in (*i*) the long-term changes in the planform of the meandering rivers; and (*ii*) the structural stability of the infrastructure installed in meandering rivers. Hence, we examined numerically predicted scour and deposition patterns to reveal the links between river geometries. As shown in Fig. 4, the scour and deposition regions follow a general trend in which the scour occurs along the outer bank, whereas the sediment deposition and point bars are located near the inner bank. However, the length and position of the scour and deposition regions along the outer and inner banks, respectively, varied significantly for different testbed rivers. For example, in single-bend Rivers 2–13, the point bars formed immediately downstream of the apex, whereas in single-bend Rivers 14–17, the point bars were located at and symmetrically positioned around the apex of the inner bank. Likewise, the deep scour regions formed along the outer banks are located either downstream from the apex (e.g., in Rivers 2–8; sharp bends in Rivers 9, 26, 41, 42, and Rivers 10–13) or symmetrically located on both sides of the apex (e.g., in Rivers 14–17, and mild bends in Rivers 18, 20–24, 39–42). A similar pattern was observed in rivers with multiple bends. For example, considering River 42, the apexes of some of the bends showed symmetrical regions of scour and deposition; however, other bends in River 42 showed asymmetrical positioning of the scour/deposition region around their apex. These patterns, that is, asymmetric and symmetric regions of deposition and regions with respect to the apex of the bends, occurred in the modeled rivers with both single and multiple bends (Fig. 4). We examined various parameters of meandering river flow and geometry, such as sinuosity, radius of curvature, and Froude number, to explore the parameters that contributed to the observed scour and deposition patterns around the apex of the bends.

The simulations suggest that the ratio of the radius of curvature to the channel width, *R/w*, is a determining factor for the scour and deposition patterns in these rivers when compared with other flow and geometrical factors presented in Table 2. As seen in this table, the *R/w* ratios of the bends across the full set of testbed rivers range from 0.5 to 6.44, with a mean of 3.6 and a standard deviation of 1.6. For river bends with *R/w* ratios of less than 2, the sediment deposition bar occurred asymmetrically around the apex, with the point bar located downstream from the apex (e.g., see Rivers 12, 13, and 26 and the sharp bends of Rivers 7–10, 41, and 42 in Fig. 4). Similarly, as shown in this figure, in bends with *R/w* < 2, the scour pattern was not symmetrically distributed around the apex, with the maximum scour depth occurring downstream from the apex of the bend.

Fig. 4 also shows that as the *R/w* ratio increased from 2 to 3, the deposition and scour patterns became more symmetrically distributed around the apex of the bends. For example, Rivers 14 and 15, with *R/w* ratios of 2 and 3, respectively, had scour and sediment deposition regions that were positioned more symmetrically around the apex of the bends than the patterns observed for bends with *R/w* < 2 in Rivers 12 and 13. As *R/w* increased from 3 to 4, the symmetry of the deposition/scour regions around the apex of the bends increased. Eventually, for the bends with *R/w* > 4, the maximum scour depth and center of the point bar are located near the apex along the outer and inner banks, respectively, with a nearly symmetrical distribution of the scour and deposition patterns around the apex of the bends. This trend of the shifting of the scour and deposition regions around the apexes can be seen in Rivers 15–17 with *R/w* ratios of 3.0, 4.0, and 5.0, respectively.

In Fig. 5, we illustrate the variation in the scour and deposition patterns as a function of the radius of curvature to the channel width ratio in some example testbed rivers. This figure plots the dynamic equilibrium bed topography of the second bend of Rivers 23, 26, and 42. The plotted bends in this figure have several curves, denoted by $R_iC_j$, marking the $j^{th}$ curve of the $i^{th}$ testbed river. The *R/w* of these curves varies between 1.68 and 6.30. The *R/w* ratios of curves $R_{26}C_5$, $R_{26}C_6$, $R_{26}C_7$, $R_{26}C_8$, $R_{42}C_6$, and $R_{42}C_8$ are less than 2 (Fig. 5; Table 2). Moreover, the point bars and scour regions developed around the apexes of these curves were asymmetrically positioned with respect to the apexes. The scour and sediment deposition patterns around curve $R_{23}C_3$, with an *R/w* ratio of 3.68, were slightly more symmetrical with respect to the apex of the curve (Fig. 5a). Furthermore, the scour and sediment deposition patterns around curves $R_{42}C_5$ and $R_{42}C_7$, which have an *R/w* ratio of 6.3, were quite symmetrical with respect to the apexes of these curves (Fig. 5c). Despite some exceptions, which we will discuss below, similar trends can be extracted from the simulation results of the other testbed rivers presented in Fig. 4.

Next, we consider the bed topography at dynamic equilibrium to examine the scour and deposition patterns as a function of the planform shapes of rivers: symmetrical, asymmetrical upstream-skewed, asymmetrical downstream-skewed, compound, and confined bends. Interestingly, the testbed rivers with all planform shapes, excluding the asymmetrical downstream-skewed meanders, seem to follow the aforementioned general rule of dependency on the *R/w* ratio for their scour/deposition patterns around the meander bend apexes. The scour/deposition patterns around the apexes of the testbed rivers with symmetrical (i.e., Rivers 2, 3, 12–17, symmetrical bends of Rivers 23, 27–30, 33–37, and 40 in Fig. 4), asymmetrical upstream-skewed (i.e., Rivers 4, 20–22, and some bends of Rivers 24, 27, 28, 31, 32, 34, and 38–41 in Fig. 4), compound (i.e., Rivers 6–10, 26, and the first bend of River 41 in Fig. 4), and confined (i.e., Rivers 18 and 42 in Fig. 4) bends can be determined based on the $R/w$ ratios of these bends. For bends with $R/w < 2$, the scour patterns and point bars were consistently located downstream of the apexes. As the $R/w$ ratios of the bends increase, the scour patterns and point bars shift upstream, positioning them more symmetrically around the apexes, with near-complete symmetry occurring when $R/w$ rations are greater than 4. However, the scour and deposition patterns around the apex of the bends with asymmetrical downstream-skewed bends deviate from the aforementioned general rule of $R/w$ ratio in two ways.

First, the positioning of the scour patterns and point bars around the sharp apexes of the multiple asymmetrical downstream-skewed bends with an $R/w$ ratio of 2.64 is more symmetrical than those of other types of multiple bends with similar ranges of $R/w$ ratios. To better illustrate this observation, in Fig. 6, we plot the third bends of Rivers 24 and 25, which correspond to asymmetrical upstream-skewed and downstream-skewed planforms, respectively, with the same $R/w$ ratios of 2.64–5.8. As seen in this figure, the scour pattern and the sediment deposition region around curve $R_{24}C_{10}$ of the asymmetrical upstream-skewed bend with an $R/w$ ratio of 2.64 (Fig. 26a) is not nearly as uniformly distributed around the apex as that of curve $R_{25}C_{10}$ of the asymmetrical downstream-skewed bend (with the same $R/w$ ratio of 2.64) (Fig. 26b). We argue that such scour patterns can influence the direction of meander bend lateral migration (Durkin et al., 2017), as well as the stability of infrastructure foundations placed near or at the apex of meandering rivers (Le et al., 2018; Khosronejad et al., 2018; Khosronejad et al., 2019a).

Second, our simulation results for the asymmetrical downstream-skewed bends suggest that these types of bends tend to modulate the near-bed turbulent flow such that the scour and deposition patterns of the meander curves immediately downstream deviate from the general patterns observed for other types of bends. More specifically, the general rule we observed for scour and deposition patterns in other meandering beds suggests that if the $R/w$ ratio is greater than four, the scour and sediment deposition around the apex of the curve are roughly symmetrical. For example, this general pattern can be observed in Fig. 6a for curve $R_{24}C_{11}$ in River 24, which consists of multiple asymmetrical upstream-skewed bends. In Fig. 6b, the meandering curve immediately downstream of the skewed curve $R_{24}C_{10}$ is curve $R_{24}C_{11}$, with an $R/w$ ratio of 5.8.

Consistent with the stated general scour/deposition rule for curves with *R/w* > 4, the scour and deposition around the apex of curve $R_{24}C_{11}$ is almost uniformly distributed around its apex. In contrast, as seen in Fig. 6b, curve $R_{25}C_{11}$, with *an R/w* ratio of 5.8, markedly deviates from the general scour and deposition pattern. In addition, only minor scour and sediment deposition occurred near the outer bank at the apex of curve $R_{25}C_{11}$. We argue that such different morphodynamic behaviors at curve $R_{25}C_{11}$ can be attributed to the modification of the flow field at the curve immediately upstream, that is, the asymmetrical downstream-skewed curve $R_{25}C_{10}$. Similar morphodynamic patterns can be seen in other testbed rivers that contain asymmetrical downstream-skewed bends.

Finally, coupled simulations, with a morphodynamic resolution of approximately 1 m in the spanwise and longitudinal directions, captured the bedforms that developed in these testbed rivers. These bedforms can be seen in Figs. 4–6, and our analysis showed that these bedforms can be categorized as dunes and are about 0.1 to 0.4 m high and 1 to 25 m long. Given the flow and sediment material characteristics of testbed rivers, the type and dimensions of these captured bedforms are in good agreement with those reported by van Rijn (1993). Specifically, according to van Rijn's empirical equations (van Rijn, 1993) for the wavelength and amplitude of dunes, the bedforms in testbed rivers with a mean flow depth of 3.3 m, bed slope of 0.0016, and median particle diameter of 1.0 mm should be approximately 0.4 m high and 24 m long, which are within the range of the dimensions of the numerically captured bedforms in this study.

**5.2. Bed shear stress distribution**

In this section, we present the computed bed shear stress distribution in the testbed rivers, measured under the condition of dynamic equilibrium in terms of bed bathymetry. The simulated maximum bed shear stress in the testbed rivers ranged from approximately 3 to 8 Pa. However, the local variation of the bed shear stress at dynamic equilibrium ranges from near zero to each river's maximum value, that is, approximately 3 to 8 Pa. We investigated the distribution of bed shear stress ($\tau_b$) in terms of the non-dimensional Shields parameter:

$$\theta = \frac{\tau_b}{(\rho_s-\rho)gD_{50}} \tag{10}$$

where $\rho_s$ is the density of the sediment material (= 2650 kg/m³ for quartz sand) and *g* is the gravitational acceleration. Given the range of the maximum bed shear stress in these rivers, their maximum Shields parameters range from approximately 0.2 to 0.5. Additionally, since the median grain size of bed material in all rivers is 1.0 mm, the critical bed shear stress and Shields parameter of the testbed rivers can be calculated to be 0.53 Pa and 0.03, respectively. This critical shear stress value corresponds to flat bed conditions and is slightly smaller when corrected for the slope of the bed surfaces. Therefore, bed changes are expected to occur at locations along rivers where the local Shields parameter is greater than the local critical value.

In Fig. 7, the instantaneous contours (color maps) of the Shields parameter are projected over the deformed bed topographies of the 42 testbed rivers in the dynamic equilibrium state. The Shields parameter distributions clearly indicate the turbulent bankfull flow fields over the deformed bed of the rivers. The size of the largest vortical flow structures, which were captured by the LES model, varied between the mean flow depth (h) and width (w) of the testbed rivers. As these large-scale, coherent flow structures pass through the rivers, their imprints over the deformed bed of the rivers can be observed through the regions of high Shields parameters (see the red finger-like features in Fig. 7). The snapshots of the Shields parameter distribution vary significantly with time, owing to the relatively rapid transport of the large-scale vortical structures (by the mean flow). To obtain the mean shear stress distributions of the testbed rivers, we conducted a limited time averaging of the bed shear stress in each testbed river. The limited time-averaging refers to a computational procedure in which the flow field quantities are time averaged over a relatively short period of time during which the bed geometry of the river is considered to stay frozen. The limited time-averaging was continued for a relatively small time window (i.e., 1 min), during which dynamic equilibrium bed topographies were assumed, allowing us to obtain contours of the mean Shields parameter for each river (Fig. 8). The contours of the mean Shields parameter provide representative bed shear stress distributions of the studied testbed rivers that can be utilized to gain insights into the connections between bed deformations and bed shear stress distributions. The bed shear stress data files of the testbed rivers in Figs. 6 and 7 can be downloaded from the Zenodo repository using the link provided in the Data Availability Statement.

As shown in Fig. 8, the regions with high bed shear stress were aligned with the curves of the meanders and were located close to the inner banks at the apexes. However, a close examination of the Shields parameter contours reveals that the high shear stress regions are slightly away from the riverbank at the apexes, owing to the no-slip boundary conditions over the rigid surfaces of the banks. In other words, the bed shear stresses are the highest over the mobile bed, slightly removed from the inner bank curves at the apexes. Moreover, immediately downstream from the apexes, the high bed shear stress regions begin to move away from the inner-bank curves and are positioned closer to the centerlines and even outer-bank curves of the rivers. These observations are consistent with those reported by Kashyap et al. (2012) for rigid flatbed rivers. They reported that "For the high curvature flatbed channels ($R/w$ less than or equal to 3), the region of highest bed shear stress was located near the inner bank close to the bend entrance, and gradually moved toward the outer bank." In rivers with multiple bends (Rivers 23–42) or single-bend rivers with multiple curves (Rivers 1–11 and 18–22), this shift in the position of high bed shear stress regions continues until eventually the high shear stress regions are aligned with and near the inner-bank curve of the next apex. On the other hand, in testbed rivers with a single bend and curve (i.e., Rivers 12–17), the shift of high shear stress regions downstream from the apexes occurs more vividly, such that the high bed shear stress regions are positioned very close to the outer bank curves. These observations regarding the bed shear stress distribution can be made for all the studied testbed rivers, regardless of their geometrical and/or hydrodynamic characteristics.

For example, in Fig. 9, we illustrate this observed trend for the bed shear stress distribution in a symmetric bend (River 14, Fig. 9b), asymmetric upstream-skewed bend (third bend of River 24, Fig. 9c), asymmetric downstream-skewed bend (third bend of River 30, Fig. 9d), compound bend (River 10, Fig. 9a), and confined meander bend (third bend of River 42, Fig. 9e). The ridges of the high bed shear stress are marked with dashed lines, which illustrate the lateral shifting of the high shear stress region in these testbed rivers. The lateral position of these dashed lines along the rivers follows the above-mentioned pattern; that is, they move closer to the inner-bank curves at the apex location and shift laterally toward the opposite riverbanks farther downstream of the apexes. Interestingly, while the overall patterns of the bed shear stress vary with the planform shape, they are relatively insensitive to other geometrical (e.g., $R/w$ ratio and sinuosity) and hydrodynamic characteristics.

Although the spatial distributions of the bed shear stress and lateral positioning of the high shear stress regions within these meanders seem to be relatively independent of the geometrical and hydrodynamic characteristics of the rivers, the magnitude of the bed shear stress was found to be a function of the $R/w$ ratio, bed slope, and flow field characteristics. For example, as shown in Table 3, the bed slope of River 19 is greater than those of other testbed rivers, which has resulted in an increase in the magnitude of the bed shear stress in this river, compared to other geometrically similar meanders (e.g., River 24) (see Fig. 8). River 20, on the other hand, has a relatively lower bed shear stress due to its relatively small bed slope (Fig. 8). River 22, which has the highest mean flow depth (= 10 m), has relatively high bed shear stresses compared to geometrically similar meanders, such as River 21 (Fig. 8).

The bed slope and flow field effects aside, the simulations suggest a relationship between the *R/w* ratio and magnitude of the bed shear stress in these testbed rivers. For example, River 12, which has a relatively small *R/w* ratio of 2.0, has elevated bed shear stress values near the apex when compared with Rivers 13–17 (Fig. 8). More specifically, in the single symmetrical Rivers 12–17, as the *R/w* ratio increases from 2.0 to 5.0, the magnitude of the bed shear stress decreases (Fig. 8). Likewise, the magnitude of the bed shear stress at the apex of bends with a smaller *R/w* ratio is somewhat greater than that of bends with a greater *R/w* ratio. For example, we can consider Rivers 10 and 24 in Figs. 9a and 9c, respectively. For these cases, the mean Shields parameter at curve $R_{10}C_3$ (with an *R/w* ratio of 1.98) was approximately 30% and 14% greater than the mean Shields parameters at curves $R_{10}C_1$ (with an *R/w* ratio of 4.93) and $R_{10}C_2$ (with an *R/w* ratio of 3.91), respectively. Additionally, the mean Shields parameters at curves $R_{24}C_{10}$ and $R_{24}C_{12}$ (both with an *R/w* ratio of 2.64) were approximately 32% greater than those at curve $R_{24}C_{11}$ (with an *R/w* ratio of 5.8).

It is important to note that our findings for the bed shear stress distributions and bed deformations in these testbed rivers contrast with the general understanding of the connection between shear stress, scour, and sediment deposition regions. It is generally thought that, in a typical meandering river, regions with high bed shear stress correspond to regions of deep scour, whereas point bars are expected to form in regions where the bed shear stress is low. However, the present simulation results suggest that point bars consistently form at the apex near the inner banks, where the bed shear stress is at its maximum. In addition, our simulation results for regions with relatively low bed shear stress (i.e., at the apex near the outer banks) coincide with the scour regions (see Figs. 4 and 8). The simulation results of our coupled flow and morphodynamics concerning the predicted bathymetric features can be justified by a careful analysis of the key components of sediment flux over the surface of the mobile bed. Furthermore, as described in Eq. 4, these components include (*i*) the velocity vector of sediment particles, $\mathbf{u}_{BL}$, and (*ii*) the sediment concentration field at the interface of water and sediment, $C_{BL}$. We argue that an increase in bed shear stress could lead to a high sediment concentration at the water/sediment interface. However, the velocity vector of the bed material determines where the mobilized bed material is transported.

Therefore, it is important to examine the streamlines of sediment particles to determine the destination of the transported bed material and, consequently, the understand the resulting bathymetry of the various river beds.

For example, we illustrate how the computed velocity vector and concentration field of the bed material result in the formation of scour regions near the outer bank of the meanders, where bed shear stress is relatively low. For the sake of brevity, let us concentrate on the simulation results for the second bend of River 42. The instantaneous and mean bed shear stress distributions of the meander are shown in Figs. 7 and 8, respectively. In addition, Fig. 9e depicts the mean bed shear stress distribution at the second bend of the river. As shown in this figure, the ridge of the high shear stress at the apex of curve $R_{42}C_7$ is close to the inner bank. In contrast, the region near the apex of the outer bank of this curve corresponds to a relatively low bed shear stress. Consequently, the sediment concentration $C_{BL}$ at the inner bank was greater than that at the outer bank.

To better illustrate the relationship between bed shear stress and sediment concentration, we plotted the instantaneous sediment concentration of the second bend of River 42 in Fig. 10. In this figure, we also plot the time variation of the Shields parameter at three points, $P_1$ to $P_3$, at the water/sediment interface near the outer bank of curve $R_{42}C_7$. The points $P_1$, $P_2$, and $P_3$ are 0.62, 5.0, and 10.0 m away from the outer bank of curve $R_{42}C_7$. While the Shields parameters of points $P_2$ and $P_3$ were consistently greater than the critical value, the point closest to the outer bank (i.e., $P_1$) experienced the least bed shear stress. However, as shown in Fig. 10c, although the bed shear stresses at point $P_1$ were small, they occasionally exceeded the critical bed shear stress and produced sediment fluxes. Furthermore, in Fig. 10b, we plot the trajectories of the sediment particles (particles' paths) over the deformed geometry of the bed that to show the computed direction of sediment fluxes in this bend. As seen, the streamlines that originate from the outer bank are all directed toward the middle of the channel, as can be seen from the dashed arrow indicating the direction of particle transport. Consequently, the sediment particles, which are picked up from the region close to the outer bank, are transported toward the inner bank of the next curve, thus creating a scour region at the outer bank of curve $R_{42}C_7$ while generating a point bar at the inner bank of curve $R_{42}C_8$. *Thus, we argue that it is the coupling between the velocity and sediment concentration field which leads to the calculated morphodynamics, and that the map of* *bed shear stress alone cannot determine the shape and geometry of the deformed mobile beds of meanders.*

**Conclusions**

We conducted coupled hydrodynamics and morphodynamics simulations in 42 large-scale natural-like testbed meandering rivers under bankfull flow conditions. The hydrodynamics of the bankfull flows were resolved using large-eddy simulation. The LES model was coupled with the morphodynamics module, in which we considered both bedload and suspended sediment load. The numerical simulations were run until the bed topography of the testbed rivers was in dynamic equilibrium, in which bedforms were present. In large-scale rivers, such as the testbed rivers in this study, coupled flow and sediment dynamics simulations can be computationally expensive. To alleviate the high computational cost of the simulations without jeopardizing the accuracy of the results, we employed our dual time-step approach and desynchronized the computations of the flow and sediment dynamics by considering morphodynamic time steps that were approximately 500 times greater than those of the flow field computations. We also ran simulations for idealized conditions using a single grain size of non-cohesive sediments and non-deformable riverbanks.

The cross-sectional geometry, planform geometry, and hydraulic conditions of the 42 numerical testbed rivers were designed to mimic the hydro- and morphodynamics conditions in natural meanders. The geometries of the testbed rivers (in MATLAB ® format) are made available via a Zenodo repository link in the data availability section. Specifically, these rivers were designed to systematically vary the bend shape, radius of curvature relative to the channel width ($R/w$), channel aspect ratio ($w/h$), relative roughness ($h/D_{50}$), and order of meander bend shapes along the channel. As a result, the testbed meandering rivers range from 0.6–15.5 km long, 33–300 m wide, and 1–10 m deep, and include various meander bend shapes, planforms, and hydrodynamics to represent natural meandering rivers as much as possible. The $R/w$ ratio of the rivers varies between 0.5 and 6.3, and the mean-flow velocities of the selected testbed rivers, under the bankfull conditions, range from 1.11 to 1.96 m/s. The selected rivers were of various types, such as single and multiple symmetric bends, asymmetric upstream-skewed bends, asymmetric downstream-skewed bends, compound bends, and confined meander bends.

The source file of the computed bed topography of the testbed rivers at their dynamic equilibrium (readable by Tecplot ® and Paraview ®) are provided via a Zenodo repository link in the Data Availability Statement. Analysis of the computed bed deformations at dynamic equilibrium showed that most of the meandering rivers studied followed a general trend for the scour and point bar regions. Specifically, we found that for meandering bends with $R/w < 2$, the scour and point bar regions were located downstream from the apex of the bends. As the bends' $R/w$ ratio increases, the scour and point bar regions shift upstream to a position more symmetrically arranged around the apexes. Relatively symmetrical positioning of these regions around the apexes occurred at $R/w > 4$. However, the scour and deposition patterns around the apexes of asymmetrical downstream-skewed bends deviate from this generalization. Namely, the scour and point bar regions around the sharp apexes of the multiple asymmetrical downstream-skewed bends with $R/w < 3$ were more symmetrical than those of the other types of multiple bends with similar $R/w$ ratio ranges. Additionally, our simulation results for the morphodynamics of the asymmetrical downstream-skewed bends revealed that these types of bends tend to modulate the near-bed turbulent flow such that only minor scour and sediment deposition are induced at the meander curves immediately downstream and, thus, drastically deviate from the general patterns observed for other types of bends. We demonstrated the importance of the coupling between the velocity and sediment concentration fields for quantifying the dynamic equilibrium morphodynamics of meandering rivers. Moreover, the map of the bed shear stress alone cannot determine the shape and geometry of the deformed mobile beds of the meanders.

Our simulation results for the bed shear stress distributions in the testbed rivers are presented in terms of the non-dimensional Shields parameter (readable by Tecplot ® and Paraview ®) and made available via a Zenodo repository link in the data availability section. We conducted limited time averaging to obtain the mean shear stress distributions of the testbed rivers at dynamic equilibrium. To determine a distinct pattern for the bed shear stress distributions, we examined the contours of the mean Shields parameter and visualized the ridges of high shear stress along the rivers. These analyses revealed that regions with high bed shear stress were aligned with the curves of the meanders and were located close to the inner banks at the apexes. Immediately downstream from the apexes, the regions of high bed shear stress moved away from the inner bank curves and toward the channel centerline. This shift continues until eventually the high shear stress regions align with and approach the inner-bank curve downstream of the apex. These observations for the bed shear stress distribution were valid for all the studied testbed rivers, regardless of their geometrical and/or hydrodynamic characteristics.

By analyzing the bed shear stress distribution, trajectories of the sediment particles, and sediment transport patterns in these meandering bends, we illustrated that the coupling between the flow field and sediment concentration field is imperative for adequate bed morphodynamics calculations, and that the bed shear stress distribution alone cannot correctly describe the geometry of deformed mobile beds of meandering rivers.

Finally, it is important to note that the simplifying assumptions in these simulations could be relaxed in future work by incorporating multiple sediment grain sizes, cohesive sediments, and deformable riverbanks. We plan to employ these model extensions in future studies.

**Data Availability Statement**

The code for the numerical model (doi:10.5281/zenodo.4677354), geometric data for the testbed rivers (doi:10.5281/zenodo.6568305), and simulation results described in this paper for the morphodynamics (doi:10.5281/zenodo.6569591) and mean bed shear stress (doi:10.5281/zenodo.6569583) are available in the Zenodo repository.

**Acknowledgments**

This work was supported by the National Science Foundation Hydrology and Geomorphology and Land-use Dynamics Program (grants EAR-0120914 and EAR-1823530). Partial support was provided by the U.S. Department of Energy's Office of Energy Efficiency and Renewable Energy (EERE) under Water Power Technologies Office (WPTO) Award No. DE-EE0009450. The views expressed herein do not necessarily represent the views of the U.S. Department of Energy or the U.S. Government. Computational resources were provided by the Civil Engineering Department, Stony Brook Research Computing and Cyberinfrastructure, and the Institute for Advanced Computational Sciences at Stony Brook University.

**Table 1.** Key characteristics of channel geometry identified in previous studies of channel migration and hydrodynamics. The latter, typically focused on radius of curvature ($R/w$) and channel aspect ratio ($w/h$), impacts turbulent flow and boundary shear stress in single, symmetric bends. We systematically isolated and varied each of these characteristics to design the meandering channel testbeds shown in Figure 2.

| Characteristic | Notes | References |
|---|---|---|
| Bend type | Simple (single lobe), compound (multiple lobes) | Brice (1974), Frothingham & Rhoads (2003), Guneralp & Marston (2012) |
| Symmetry | Symmetric, asymmetric (upstream vs. downstream skew) | Carson & Lapointe (1983); Guo et al. (2019), Abad and Garcia (2009a,b) |
| Radius of curvature ($R/w$) | Key predictor for channel migration rates; hydrodynamics typically studied for single, symmetric bends | Hickin and Nanson (1975), Howard & Knutson (1984), Blanckaert (2011), Sylvester et al. (2019) |
| Aspect ratio ($w/h$) | Hydrodynamics typically studied for single, symmetric bends | Blanckaert (2011), Kashyap et al. (2012) |

**Table 2.** Geometry and flow parameters for the testbed rivers (1 to 42). Channel parameters are length ($L$), width ($w$), depth ($h$), slope ($S$), sinuosity ($Si$), and radius of curvature relative to channel width ($R/w$). The flow parameters are mean-flow velocity ($U$) and Froude number ($Fr$). We note that the median grain size of the sediment materials of these testbed rivers is 1.0 mm and that the Reynolds numbers of these river (bankfull) flows are in the order of $O(10^6)$ or higher.

| River | L (m) | w (m) | h (m) | S | Si | U (m/s) | R/w | Fr |
|---|---|---|---|---|---|---|---|---|
| 1 | 1410 | 100 | 3.3 | 0.00015 | 1.17 | 1.52 | 2.86 | 0.27 |
| 2 | 2110 | 100 | 3.3 | 0.00016 | 1.76 | 1.75 | 2.40 | 0.31 |
| 3 | 4440 | 100 | 3.3 | 0.00016 | 3.70 | 1.75 | 3.68 | 0.31 |
| 4 | 4580 | 100 | 3.3 | 0.00016 | 3.83 | 1.75 | 2.64 – 5.80 | 0.31 |
| 5 | 4580 | 100 | 3.3 | 0.00016 | 3.83 | 1.75 | 2.64 – 5.80 | 0.31 |
| 6 | 3040 | 100 | 3.3 | 0.00016 | 1.79 | 1.75 | 2.58 | 0.31 |
| 7 | 2820 | 100 | 3.3 | 0.00016 | 2.99 | 1.75 | 1.72 | 0.31 |
| 8 | 3280 | 100 | 3.3 | 0.00016 | 4.96 | 1.75 | 1.71 | 0.31 |
| 9 | 2740 | 100 | 3.3 | 0.00016 | 3.25 | 1.75 | 1.71 – 4.24 | 0.31 |
| 10 | 3188 | 100 | 3.3 | 0.00016 | 3.25 | 1.75 | 1.98 – 4.93 | 0.31 |
| 11 | 3790 | 100 | 3.3 | 0.00016 | 2.69 | 1.75 | 1.31 – 2.95 | 0.31 |
| 12 | 600 | 100 | 3.3 | 0.00016 | 4.15 | 1.75 | 0.50 | 0.31 |
| 13 | 1000 | 100 | 3.3 | 0.00016 | 3.47 | 1.75 | 1.00 | 0.31 |
| 14 | 1820 | 100 | 3.3 | 0.00016 | 3.16 | 1.75 | 2.00 | 0.31 |
| 15 | 2640 | 100 | 3.3 | 0.00016 | 3.05 | 1.75 | 3.00 | 0.31 |
| 16 | 3460 | 100 | 3.3 | 0.00016 | 3.00 | 1.75 | 4.00 | 0.31 |
| 17 | 4260 | 100 | 3.3 | 0.00016 | 2.96 | 1.75 | 5.00 | 0.31 |
| 18 | 2930 | 100 | 3.3 | 0.00016 | 1.46 | 1.75 | 1.68 – 6.30 | 0.31 |
| 19 | 1527 | 33 | 3.3 | 0.00025 | 3.83 | 1.96 | 2.66 – 5.86 | 0.34 |
| 20 | 13740 | 300 | 3.3 | 0.00011 | 3.83 | 1.31 | 2.40 – 5.27 | 0.23 |
| 21 | 1527 | 33 | 1 | 0.0004 | 3.83 | 1.11 | 2.93 – 6.44 | 0.35 |
| 22 | 13740 | 300 | 10 | 0.00006 | 3.83 | 1.89 | 2.64 – 5.80 | 0.19 |
| 23 | 13320 | 100 | 3.3 | 0.00016 | 3.70 | 1.75 | 3.68 | 0.31 |
| 24 | 13740 | 100 | 3.3 | 0.00016 | 3.83 | 1.75 | 2.64 – 5.80 | 0.31 |
| 25 | 13740 | 100 | 3.3 | 0.00016 | 3.83 | 1.75 | 2.64 – 5.80 | 0.31 |
| 26 | 8760 | 100 | 3.3 | 0.00016 | 2.80 | 1.75 | 1.72 | 0.31 |
| 27 | 13460 | 100 | 3.3 | 0.00016 | 3.74 | 1.75 | 2.64 – 5.80 | 0.31 |
| 28 | 13600 | 100 | 3.3 | 0.00016 | 3.78 | 1.75 | 2.64 – 5.80 | 0.31 |
| 29 | 13460 | 100 | 3.3 | 0.00016 | 3.74 | 1.75 | 2.64 – 5.80 | 0.31 |
| 30 | 13600 | 100 | 3.3 | 0.00016 | 3.78 | 1.75 | 2.64 – 5.80 | 0.31 |
| 31 | 13750 | 100 | 3.3 | 0.00016 | 3.67 | 1.75 | 2.64 – 5.80 | 0.31 |
| 32 | 13600 | 100 | 3.3 | 0.00016 | 3.78 | 1.75 | 2.64 – 5.80 | 0.31 |
| 33 | 15044 | 100 | 3.3 | 0.00016 | 3.37 | 1.75 | 3.68 – 3.76 | 0.31 |
| 34 | 15487 | 100 | 3.3 | 0.00016 | 3.43 | 1.75 | 2.64 – 5.80 | 0.31 |
| 35 | 10990 | 100 | 3.3 | 0.00016 | 3.05 | 1.75 | 2.41 – 3.68 | 0.31 |
| 36 | 8660 | 100 | 3.3 | 0.00016 | 2.40 | 1.75 | 2.41 – 3.68 | 0.31 |

| 37 | 9380 | 100 | 3.3 | 0.00016 | 3.23 | 1.75 | 3.68 | 0.31 |
| 38 | 9660 | 100 | 3.3 | 0.00016 | 3.34 | 1.75 | 2.64 – 5.80 | 0.31 |
| 39 | 9520 | 100 | 3.3 | 0.00016 | 3.29 | 1.75 | 2.64 – 5.80 | 0.31 |
| 40 | 9520 | 100 | 3.3 | 0.00016 | 3.29 | 1.75 | 2.64 – 5.80 | 0.31 |
| 41 | 5361 | 100 | 3.3 | 0.00016 | 2.71 | 1.75 | 1.39 – 2.58 | 0.31 |
| 42 | 8790 | 100 | 3.3 | 0.00016 | 1.46 | 1.75 | 1.68 – 6.30 | 0.31 |

**Table 3.** The sequence of meander bends for testbeds 23 to 42. The right column denotes the order of meander bends as listed in Table 2. "Straight" indicates a straight reach separating two bends.

| Test bed | Bend sequence |
|---|---|
| 23 | 3-3-3 |
| 24 | 4-4-4 |
| 25 | 5-5-5 |
| 26 | 7-7-7 |
| 27 | 3-4-3 |
| 28 | 4-3-4 |
| 29 | 3-5-3 |
| 30 | 5-3-5 |
| 31 | 3-4-5 |
| 32 | 3-5-4 |
| 33 | 3-7-3 |
| 34 | 4-7-4 |
| 35 | 3-2-3 |
| 36 | 2-3-2 |
| 37 | 3-straight-3 |
| 38 | 4-straight-4 |
| 39 | 3-straight-4 |
| 40 | 4-straight-3 |
| 41 | 7-straight-4 |
| 42 | 18-18-18 |

**Table 4.** Computational grid resolutions and time steps of the coupled simulations for the 42 testbed rivers. $N_x$, $N_y$, and $N_z$ are the number of grid nodes to discretize the flow domain, which result in the grid resolution of $\Delta x$, $\Delta y$, and $\Delta z$ in longitudinal, spanwise, and vertical directions, respectively, and a total number of computational grid nodes $N_f$. The mobile sediment beds are discretized with $N_m$ triangular cells and a resolution of $\Delta s$. The time steps of hydrodynamics and morphodynamics computations are $\Delta t_f$ and $\Delta t_m$, respectively.

| River | $N_x \times N_y \times N_z$ | $\Delta x \times \Delta y \times \Delta z$ | $N_f \times 10^6$ | $\Delta s$ | $N_m \times 10^3$ | $\Delta t_f$ | $\Delta t_m$ |
|---|---|---|---|---|---|---|---|
| 1 | 3501×225×25 | 0.40×0.44×0.14 | 19.69 | 0.94 | 157.5 | 0.05 | 25 |
| 2 | 5001×225×25 | 0.42×0.44×0.14 | 28.13 | 0.95 | 225.0 | 0.06 | 30 |
| 3 | 10001×225×25 | 0.44×0.44×0.14 | 56.25 | 0.98 | 449.0 | 0.06 | 30 |
| 4 | 10013×225×25 | 0.46×0.44×0.14 | 56.32 | 1.00 | 450.6 | 0.06 | 30 |
| 5 | 10013×225×25 | 0.46×0.44×0.14 | 56.32 | 1.00 | 450.6 | 0.06 | 30 |
| 6 | 6753×225×25 | 0.45×0.44×0.14 | 37.99 | 0.99 | 303.88 | 0.06 | 30 |
| 7 | 6101×225×25 | 0.46×0.44×0.14 | 34.32 | 0.98 | 274.54 | 0.06 | 30 |
| 8 | 7101×225×25 | 0.46×0.44×0.14 | 39.94 | 0.99 | 319.54 | 0.06 | 30 |
| 9 | 6001×225×25 | 0.46×0.44×0.14 | 33.76 | 0.98 | 271.22 | 0.06 | 30 |
| 10 | 6901×225×25 | 0.46×0.44×0.14 | 38.82 | 0.99 | 310.99 | 0.06 | 30 |
| 11 | 8201×225×25 | 0.46×0.44×0.14 | 46.13 | 0.98 | 274.54 | 0.06 | 30 |
| 12 | 1501×225×25 | 0.40×0.44×0.14 | 8.44 | 0.93 | 67.51 | 0.06 | 30 |
| 13 | 2381×225×25 | 0.42×0.44×0.14 | 13.39 | 0.94 | 107.15 | 0.06 | 30 |
| 14 | 4301×225×25 | 0.42×0.44×0.14 | 24.19 | 0.96 | 193.51 | 0.06 | 30 |
| 15 | 5861×225×25 | 0.45×0.44×0.14 | 32.97 | 0.99 | 263.74 | 0.06 | 30 |
| 16 | 7601×225×25 | 0.46×0.44×0.14 | 42.76 | 1.00 | 342.05 | 0.06 | 30 |
| 17 | 9201×225×25 | 0.46×0.44×0.14 | 51.76 | 1.00 | 414.06 | 0.06 | 30 |
| 18 | 6381×225×25 | 0.46×0.44×0.14 | 35.89 | 1.00 | 287.15 | 0.06 | 30 |
| 19 | 3401×73×25 | 0.40×0.45×0.14 | 6.21 | 0.93 | 49.66 | 0.06 | 30 |
| 20 | 27481×601×25 | 0.50×0.50×0.14 | 412.89 | 1.12 | 3303.21 | 0.08 | 40 |
| 21 | 3401×73×13 | 0.45×0.45×0.08 | 3.23 | 0.99 | 49.65 | 0.05 | 25 |
| 22 | 27481×601×49 | 0.50×0.50×0.20 | 809.29 | 1.12 | 3303.21 | 0.07 | 35 |
| 23 | 26641×225×25 | 0.5×0.44×0.14 | 149.85 | 1.04 | 1198.84 | 0.06 | 30 |
| 24 | 27481×225×25 | 0.50×0.44×0.14 | 154.58 | 1.04 | 1236.64 | 0.06 | 30 |
| 25 | 27481×225×25 | 0.50×0.44×0.14 | 154.58 | 1.04 | 1236.64 | 0.06 | 30 |
| 26 | 19001×225×25 | 0.46×0.44×0.14 | 106.88 | 1.00 | 855.03 | 0.06 | 30 |
| 27 | 26921×225×25 | 0.50×0.44×0.14 | 151.43 | 1.04 | 1211.14 | 0.06 | 30 |
| 28 | 27201×225×25 | 0.50×0.44×0.14 | 153.00 | 1.04 | 1224.04 | 0.06 | 30 |
| 29 | 26921×225×25 | 0.46×0.44×0.14 | 106.88 | 1.00 | 1211.14 | 0.06 | 30 |
| 30 | 27201×225×25 | 0.50×0.44×0.14 | 153.00 | 1.04 | 1224.02 | 0.06 | 30 |
| 31 | 27481×225×25 | 0.50×0.44×0.14 | 154.58 | 1.04 | 1236.64 | 0.06 | 30 |
| 32 | 27201×225×25 | 0.50×0.44×0.14 | 153.00 | 1.04 | 1224.10 | 0.06 | 30 |
| 33 | 30001×225×25 | 0.50×0.44×0.14 | 168.75 | 1.04 | 1350.11 | 0.06 | 30 |
| 34 | 30901×225×25 | 0.50×0.44×0.14 | 173.82 | 1.04 | 1390.43 | 0.06 | 30 |
| 35 | 21981×225×25 | 0.50×0.44×0.14 | 123.64 | 1.04 | 989.17 | 0.06 | 30 |
| 36 | 19001×225×25 | 0.46×0.44×0.14 | 106.88 | 1.00 | 855.03 | 0.06 | 30 |
| 37 | 19001×225×25 | 0.49×0.44×0.14 | 106.88 | 1.00 | 855.03 | 0.06 | 30 |
| 38 | 19321×225×25 | 0.50×0.44×0.14 | 108.67 | 1.04 | 869.24 | 0.06 | 30 |
| 39 | 19001×225×25 | 0.50×0.44×0.14 | 106.88 | 1.04 | 855.03 | 0.06 | 30 |
| 40 | 19001×225×25 | 0.50×0.44×0.14 | 106.88 | 1.04 | 855.03 | 0.06 | 30 |
| 41 | 11601×225×25 | 0.46×0.44×0.14 | 65.26 | 1.00 | 522.06 | 0.06 | 30 |
| 42 | 19001×225×25 | 0.46×0.44×0.14 | 106.88 | 106.88 | 1.00 | 0.06 | 30 |

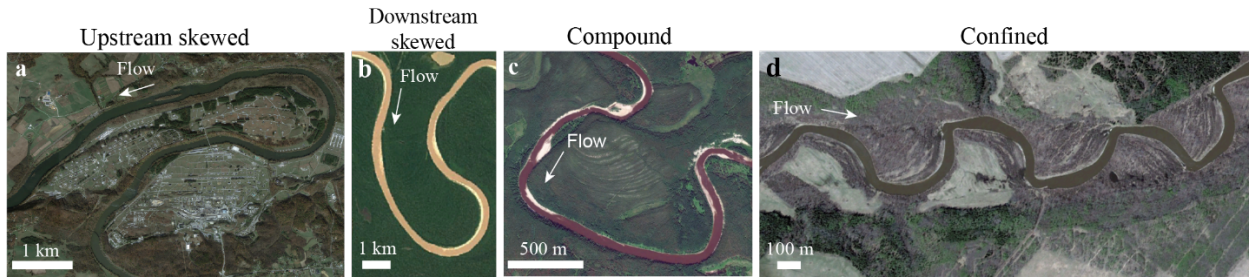

**Figure 1.** Common geometries of meandering river bends. The arrow in each panel indicates the overall flow direction. (a) An asymmetric, upstream-skewed bend on the New River at Cowan, Virginia, USA (37.19ºN, 80.55ºW); (b) An asymmetric, downstream-skewed bend on the Juruá River, Peru (6.49ºS, 68.37ºW); (c) A compound bend on the Beatton River, Canada (57.04ºN, 120.99ºW); (d) Confined meander bends on the Beaver River, Canada (54.40ºN, 110.63ºW); Images: Landsat/Copernicus/Maxar Technologies/Google Earth.

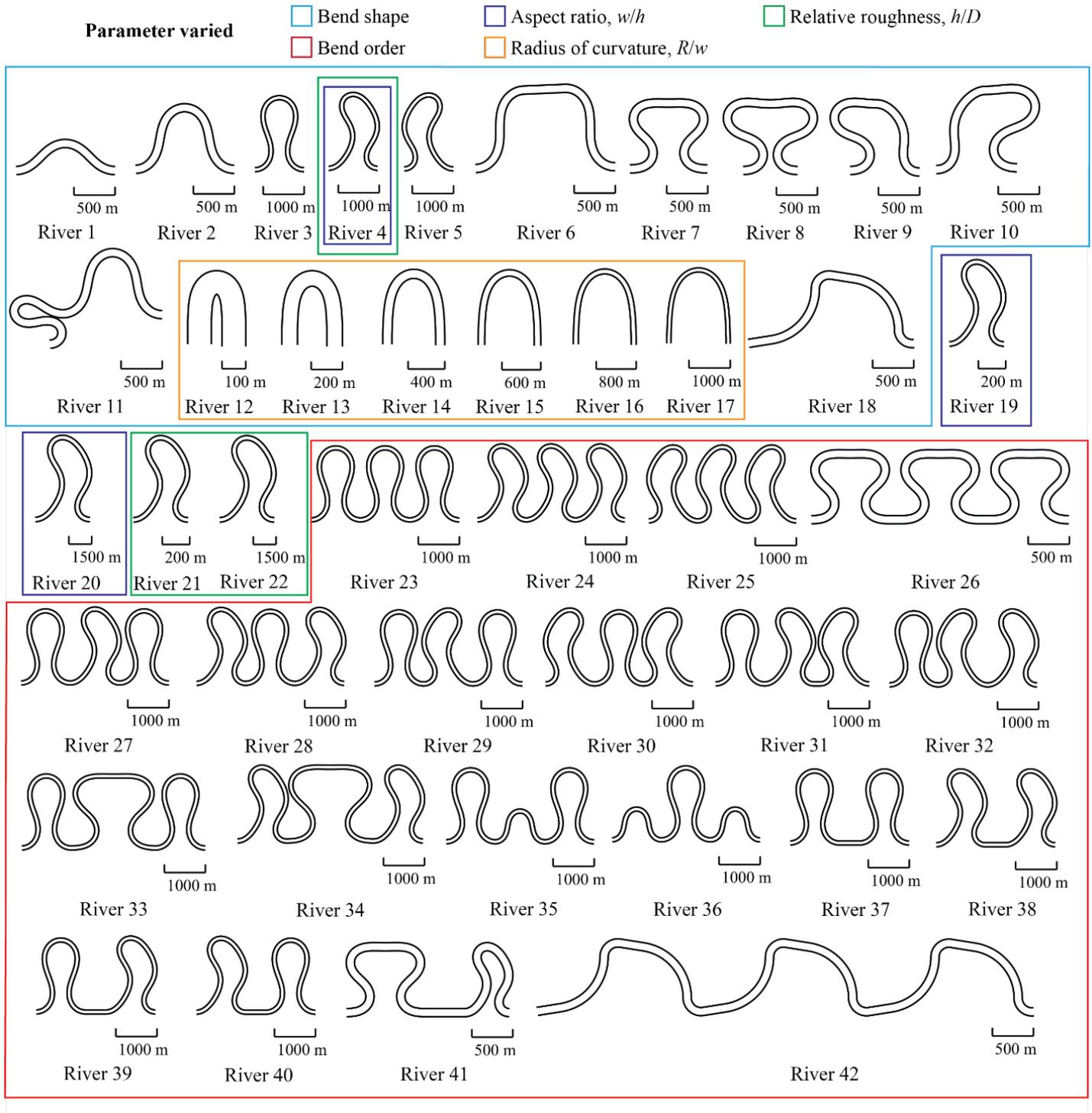

**Figure 2.** Planform geometry for the set of 42 meandering river testbeds for the coupled hydrodynamics and morphodynamics simulations. Each testbed is formed by either a single meander bend or several consecutive meander bends. The colored boxes group the testbeds according to the main independent parameter varied. For individual bends, these parameters are the bend shape (Rivers 1 to 18), radius of curvature relative to channel width ($R/w$; Rivers 12 to 17), channel aspect ratio ($w/h$; Rivers 19 and 20), and relative roughness ($h/D$; Rivers 21 and 22). The 20 remaining testbeds (Rivers 23 to 42) vary in the order of consecutive meander bends with different shapes. These channel geometries are summarized in Tables 2 and 3. The flow is from left to right.

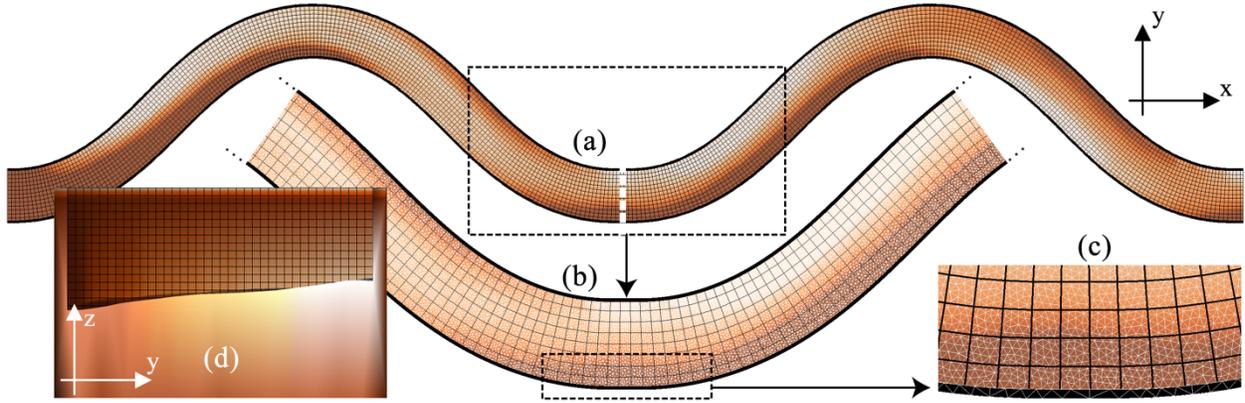

**Figure 3.** Schematic of the CURVIB method to discretize the flow and the water-sediment interface (orange color map) using the structured background (black lines) and unstructured triangular (white lines) grid systems. The riverbanks (thick black lines on the sides) and the mobile sediment bed (colored in white/brown) are discretized with unstructured triangular mesh (white lines) and treated as a sharp-interface immersed body and embedded in the background structured curvilinear grid system. (d) depicts a cross sectional view of the river taken from the white dashed line in (a). As seen in (d), the mobile bed is deformed, and the flow field is resolved on the background mesh (black lines). The brown area on the bottom of (d) is the sediment layer, which is discretized using the unstructured triangular grid system (shown from top view in (c)). For clarity, the structured grid system (black lines) is coarsened.

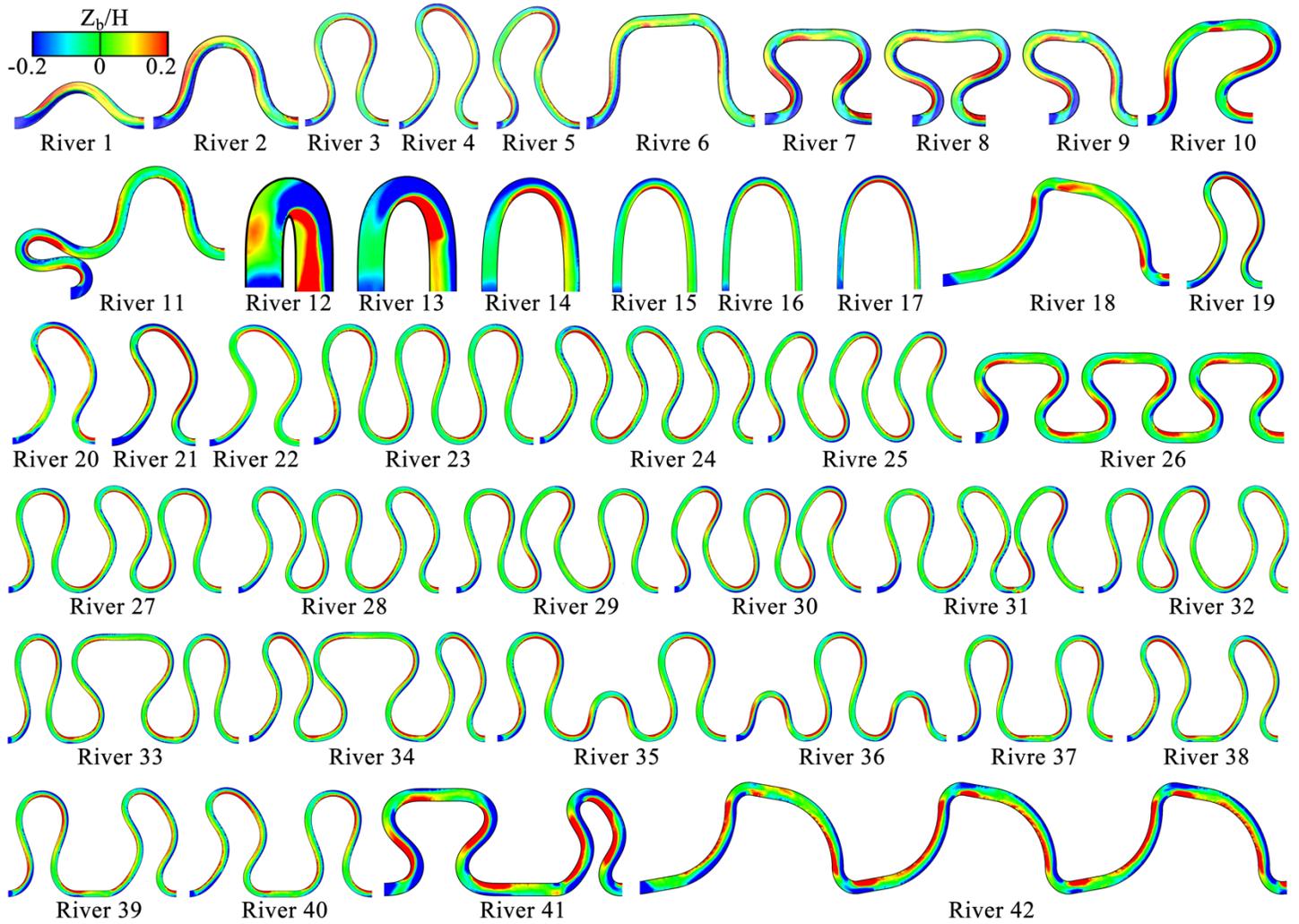

**Figure 4.** Computed bed bathymetry of the 42 meandering river testbeds at the dynamic equilibrium state. Color maps show the bed elevation ($Z_b$) of the channels at dynamic equilibrium and relative to their initial flat-bed state. Elevation values are normalized by the mean-flow depth (H) of each river, as shown in Table 3. These data for bed topography can be found on Zenodo repository via the link provided in the Data Availability Section. The flow is from left to right.

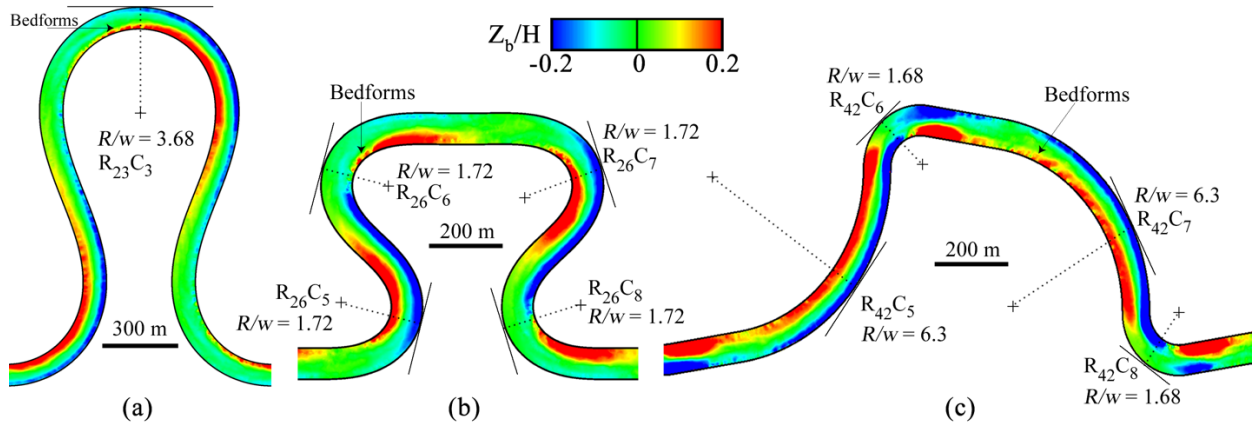

**Figure 5.** Computed bed bathymetry at dynamic equilibrium for the second bend of the testbeds for (a) River 23, (b) River 26, and (c) River 42 (c). Color maps show the bed elevation ($Z_b$) of the channels relative to their initial flat-bed state, and elevation is normalized by the mean-flow depth (H) of each river. This figure illustrates the variation of scour/deposition patterns as a function of radius of curvature (R) to the channel width (W) ratio. $R_iC_j$ represents the $j^{th}$ curve of River i. The flow is from left to right.

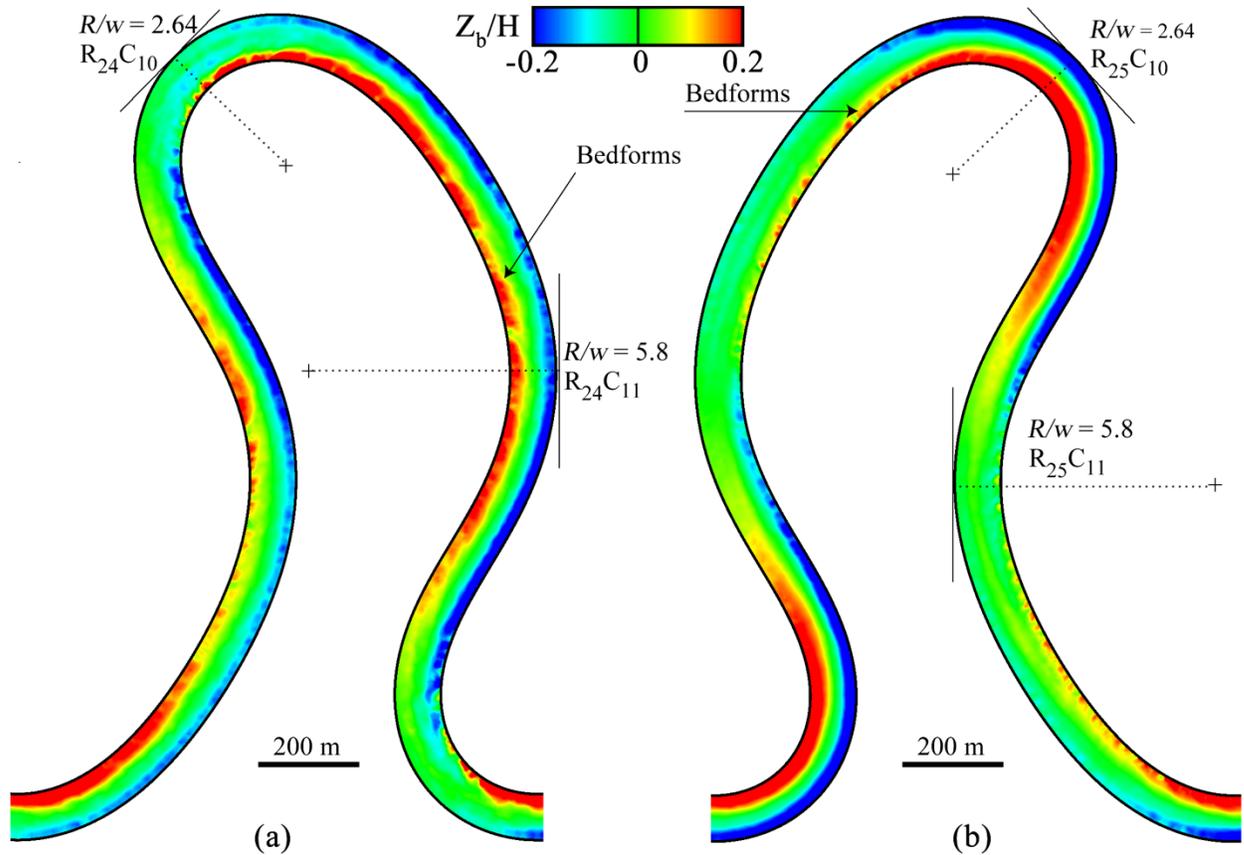

**Figure 6.** Computed bed bathymetry at dynamic equilibrium for the second bend of the testbeds for (a) River 24, which skews upstream, and (b) River 25, which skews downstream. Color maps show the bed elevation ($Z_b$) of the channels relative to their initial flat-bed state, normalized with the mean-flow depth (H) of each river. This figure illustrates the effect of the skew direction on the scour/deposition patterns for asymmetric bends. $R_iC_j$ represents the $j^{th}$ curve of River i. The flow is from left to right.

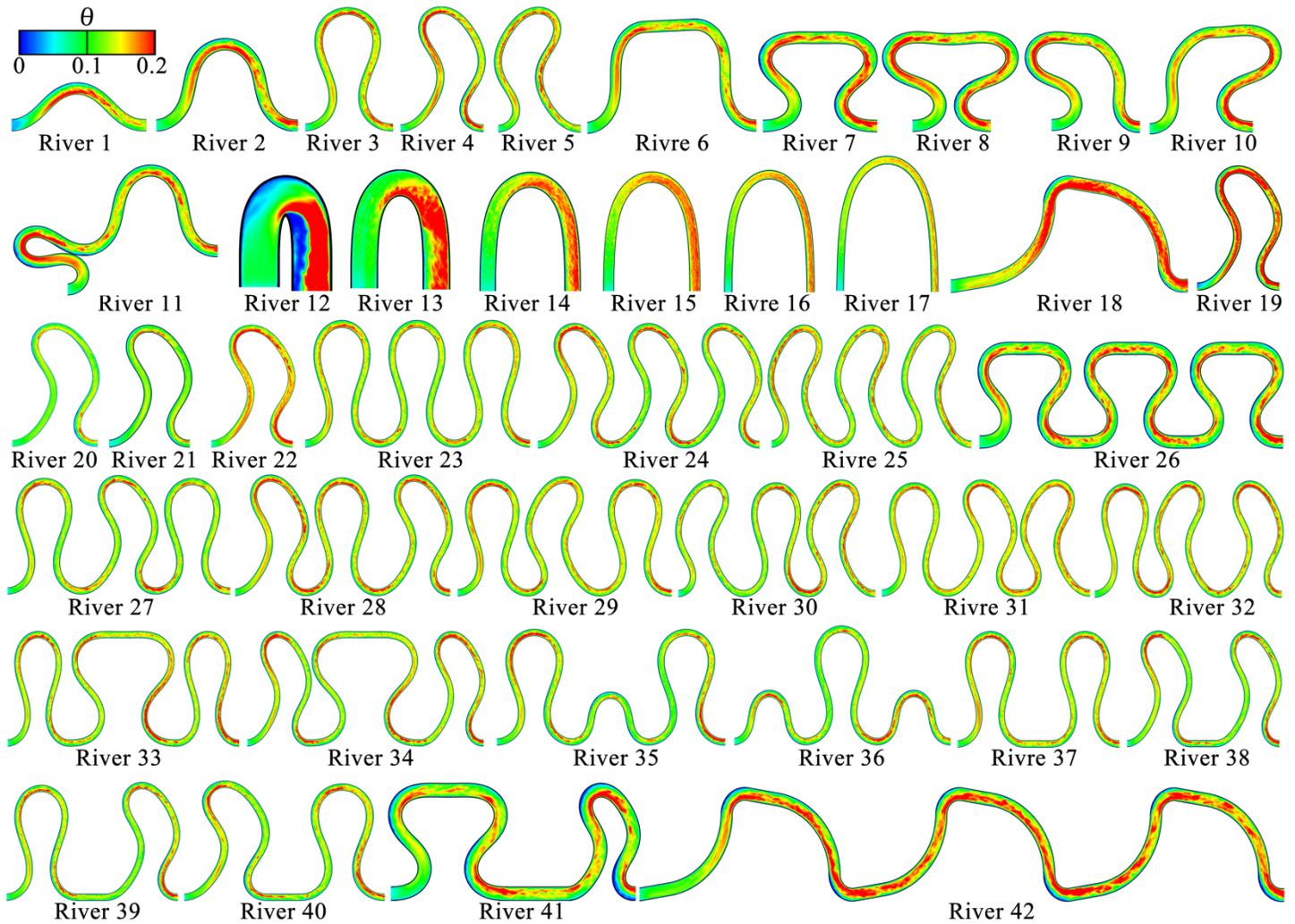

**Figure 7.** Color maps of instantaneous values of the Shields parameter, $\theta$, projected over the deformed bed of the 42 meandering testbed rivers at the dynamic equilibrium state. The data files for these non-dimensional bed shear stress distributions can be found on Zenodo repository via the link provided in the Data Availability Section. The flow is from left to right.

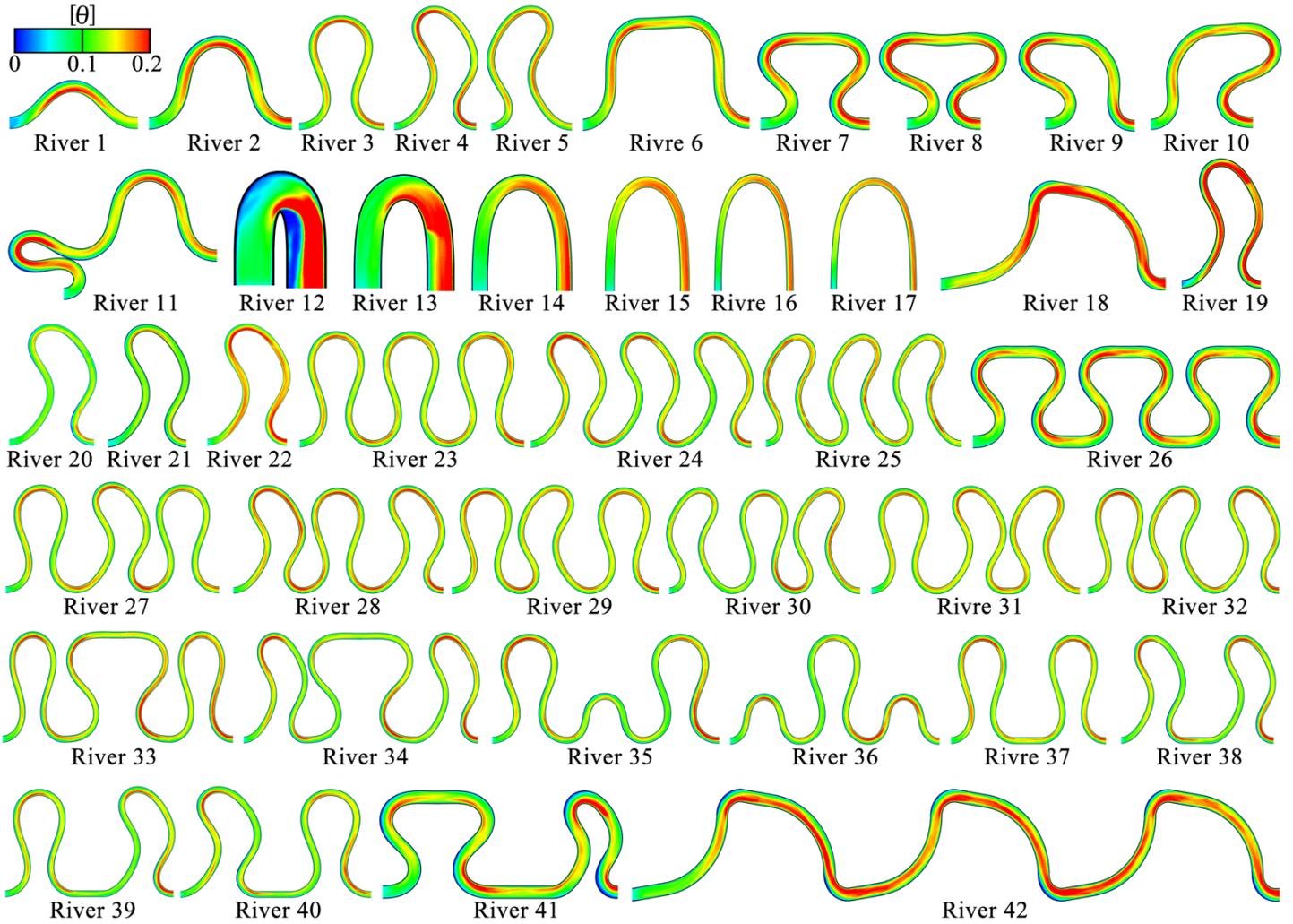

**Figure 8.** Color maps of the limited time-averaged Shields parameter, $[\theta]$, which represents the non-dimensional mean bed shear stress. Values of the Shields number are projected over the deformed bed of the 42 meandering testbed rivers at the dynamic equilibrium state. The corresponding data files can be found on Zenodo repository via the link provided in the Data Availability Section. The flow is from left to right.

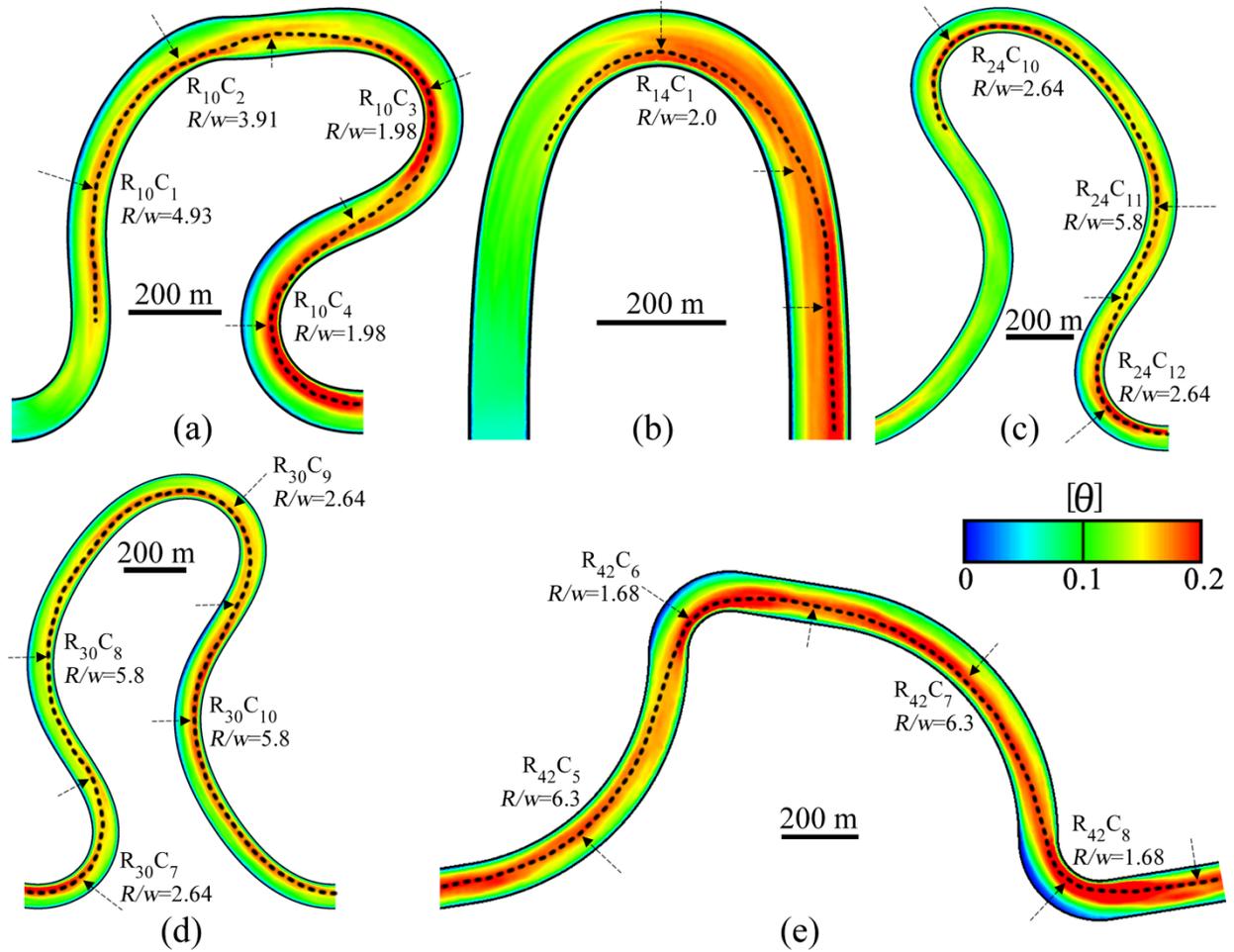

**Figure 9.** Color maps of the time-averaged Shields parameter, [θ], for a limited averaging time. Shields numbers are shown for the dynamic equilibrium state projected over the deformed bed of Rivers (a) 10, (b) 14, (c) 24 (third bend), (d) 30 (third bend), and (e) 42 (second bend). The dashed lines in (b) to (e) mark the ridges of high bed shear stress (mean values) and the dashed arrows show the spanwise shifting of the position of the ridges of high bed shear stress. As seen, regardless of the *R/w* ratio of the curves, the dashed lines are positioned near the apexes of the inner banks. Immediately downstream from the apexes, the ridges shift away from the inner banks. $R_iC_j$ represents the j$^{th}$ curve of River i. The flow is from left to right.

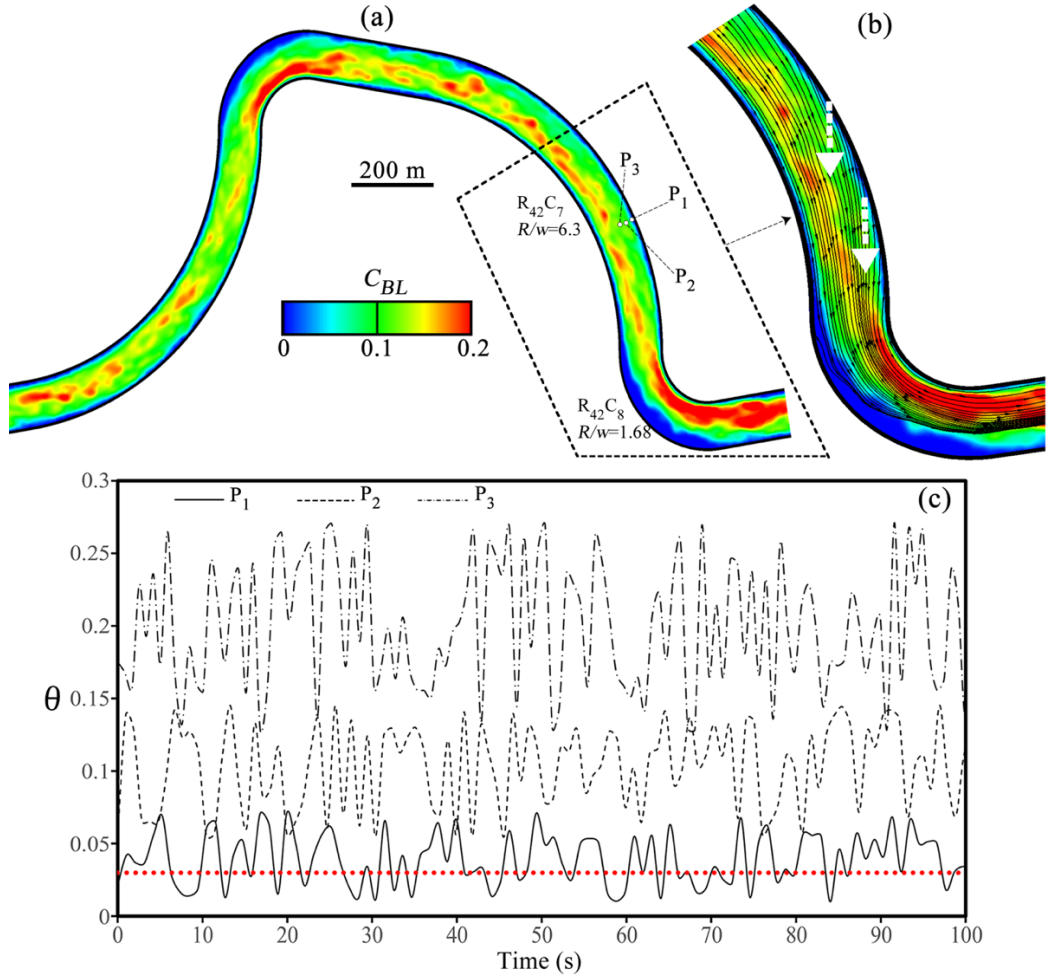

**Figure 10.** Instantaneous sediment concentration and sediment particle streamlines in the second bend of River 42. (a) Contours of instantaneous sediment concentration at the interface of the water/sediment (b) Zoomed-in view of curves $R_{42}C_7$ and $R_{42}C_8$ in which the instantaneous contours of sediment concentration are superimposed by the instantaneous streamlines of the sediment particles at the edge of the bed-load layer. The white dashed arrows in (b) mark the direction of the sediment transport, i.e., away from the outer bank at the apex of curve $R_{42}C_7$. (c) Time variation of the Shields parameter at three representative points, $P_1$ to $P_3$, which are 0.62, 5.0, and 10.0 m away from the outer bank of curve $R_{42}C_7$. The approximated positions of these points are shown in (a). The red dotted line in (c) shows the critical value of Shields parameter (0.03) for the sediment bed material. As described in Section 2.2, $C_{BL}$ is the sediment concentration at the interface of water and sediment (i.e., the edge of the bed-load layer). In (a) and (b), the flow is from left to right.